\documentclass{sig-alternate}
\usepackage{times}
\usepackage{graphicx,graphics,epsfig,subfig}

\usepackage{dsfont,comment,colortbl,cite}
\usepackage{algorithm}

\usepackage{txfonts}

\ifodd 0
\newcommand{\rev}[1]{\textcolor{blue}{#1}} 
\newcommand{\com}[1]{\textbf{\color{blue} (COMMENT: #1)}} 
\else
\newcommand{\rev}[1]{#1}
\newcommand{\com}[1]{}
\fi

\newcommand{\lyxdot}{.}

\newcommand{\beq}   {\begin{equation}}
\newcommand{\eeq}   {\end{equation}}
\newcommand{\bea}   {\begin{eqnarray}}
\newcommand{\eea}   {\end{eqnarray}}
\newcommand{\bda}   {\begin{eqnarray*}}
\newcommand{\eda}   {\end{eqnarray*}}
\newcommand{\bdalign}   {\begin{align*}}
\newcommand{\edalign}   {\end{align*}}

\newtheorem{thm}{\bf Theorem}

\newtheorem{cor}{\bf Corollary}

\newtheorem{prop}{\bf Proposition}

\newtheorem{defn}{\bf Definition}

\conferenceinfo{ACM Multimedia}{'11 Scottsdale, Arizona, USA}

\title{
Celerity: A Low-Delay Multi-Party Conferencing Solution
}

\numberofauthors{6}
\author{
\alignauthor
Xiangwen Chen\\
        \affaddr{Dept. of Information Engineering}\\
       \affaddr{The Chinese University of Hong Kong}\\
\alignauthor
Minghua Chen\\
        \affaddr{Dept. of Information Engineering}\\
       \affaddr{The Chinese University of Hong Kong}\\
\alignauthor
Baochun Li\\
       \affaddr{Dept. of Electrical and Computer Engineering}\\
       \affaddr{University of Toronto}\\
\and  
\alignauthor
Yao Zhao\\
       \affaddr{Alcatel-Lucent}\\
\alignauthor
Yunnan Wu\\
       \affaddr{Facebook Inc.}\\
\alignauthor
Jin Li \\
       \affaddr{Microsoft Research at Redmond}\\
}

\begin{document}
\maketitle

\begin{abstract}
In this paper, we attempt to revisit the problem of
multi-party conferencing from a practical perspective, and to
rethink the design space involved in this problem. We believe that an
emphasis on low end-to-end delays between any two parties in the
conference is a must, and the source sending rate in a session should
adapt to bandwidth availability and congestion. We present {\em Celerity}, a multi-party conferencing solution specifically designed to achieve our objectives.
It is entirely Peer-to-Peer (P2P), and as such eliminating the cost of
maintaining centrally administered servers. It is designed to deliver
video with low end-to-end delays, at quality levels commensurate with available network resources over arbitrary network topologies where \emph{bottlenecks can be anywhere in the network}. This is in contrast to commonly assumed P2P scenarios where bandwidth bottlenecks reside only at the edge of the network.
The highlight in our design is a distributed and adaptive rate control protocol, that can discover and adapt to
arbitrary topologies and network conditions quickly, converging to
efficient link rate allocations allowed by the underlying network. In
accordance with adaptive link rate control, source video encoding rates
are also dynamically controlled to optimize video
quality in arbitrary and unpredictable network conditions. We have
implemented {\em Celerity} in a prototype system, and demonstrate its
superior performance over existing solutions in a local experimental testbed and over the Internet.
\end{abstract}

\section{Introduction}

\label{sec:intro}

With the availability of front-facing cameras in high-end smartphone
devices (such as the Samsung Galaxy S and the iPhone 4), notebook
computers, and HDTVs, {\em multi-party} video conferencing, which
involves more than two participants in a live conferencing session,
has attracted a significant amount of interest from the industry.
Skype, for example, has recently launched a monthly-paid service supporting
multi-party video conferencing in its latest version (Skype 5) \cite{skype5}.
Skype video conferencing has also been recently supported in a range
of new Skype-enabled televisions, such as the Panasonic VIERA series,
so that full-screen high-definition video conferencing can be enjoyed
in one's living room. Moreover, Google has supported multi-party video
conferencing in its latest social network service \textit{Google+}.
And Facebook cooperating with Skype plans to provide video conferencing
service to its billions of users. We argue that these new conferencing
solutions have the potential to provide an immersive human-to-human
communication experience among remote participants. Such an argument
has been corroborated by many industry leaders: Cisco predicts that
video conferencing and tele-presence traffic will increase ten-fold
between 2008-2013 \cite{cisco}.

While traffic flows in a live multi-party conferencing session are
fundamentally represented by a multi-way communication process, today's
design of multi-party video conferencing systems are engineered in
practice by composing communication primitives ({\em e.g.,} transport
protocols) over uni-directional feed-forward links, with primitive
feedback mechanisms such as various forms of acknowledgments in TCP
variants or custom UDP-based protocols. We believe that a high-quality
protocol design must harness the full potential of the multi-way communication
paradigm, and must guarantee the stringent requirements of low end-to-end
delays, with the highest possible source coding rates that can be
supported by dynamic network conditions over the Internet.

From the industry perspective, known designs of commercially available
multi-party conferencing solutions are either largely server-based,
e.g., Microsoft Office Communicator, or are separated into multiple
point-to-point sessions (this approach is called Simulcast), e.g.,
Apple iChat. Server-based solutions are susceptible to central resource
bottlenecks, and as such scalability becomes a main concern when multiple
conferences are to be supported concurrently. In the Simulcast approach,
each user splits its uplink bandwidth equally among all receivers
and streams to each receiver separately. Though simple to implement,
Simulcast suffers from poor quality of service. Specifically, peers
with low upload capacity are forced to use a low video rate that degrades
the overall experience of the other peers.

In the academic literature, there are recently several studies on
peer-to-peer (P2P) video conferencing from a utility maximization
perspective \cite{all:Mutualcast:LPZ05,chen2008ump,akku2011peer,ponec2009multi,ponec2011multi,liangoptimal}.
Among them, Li {\em et al.} \cite{all:Mutualcast:LPZ05} and Chen
{\em et al.}~\cite{chen2008ump} may be the most related ones
to this work (we call their unified approach Mutualcast). They have
tried to support content distribution and multi-party video conferencing
in multicast sessions, by maximizing aggregate application-specific
utility and the utilization of node uplink bandwidth in P2P networks.
Specific depth-1 and depth-2 tree topologies have been constructed
using tree packing, and rate control was performed in each of the
tree-based one-to-many sessions.However, they only considered the
limited scenario where bandwidth bottlenecks reside at the edge of
the network, while in practice bandwidth bottlenecks can easily reside
in the core of the network \cite{akella2003empirical,hu2004locating}.
Further, all existing industrial and academic solutions, including
Mutualcast, did not explicitly consider bounded delay in designs,
and can lead to unsatisfied interactive conferencing experience.

\subsection{Contribution}

In this paper, we reconsider the design space in multi-party video
conferencing solutions, and present {\em Celerity}, a new multi-party
conferencing solution specifically designed to maintain low end-to-end
delays while maximizing source coding rates in a session. {\em Celerity}
has the following salient features:
\begin{itemize}
\item It operates in a pure P2P manner, and as such eliminating the cost
of maintaining centrally administered servers.
\item It can deliver video at quality levels commensurate with available
network resources over \emph{arbitrary network topologies}, while
maintaining \textit{bounded end-to-end delays}.
\item It can automatically adapt to unpredictable network dynamics, such
as cross traffic and abrupt link failures, allowing smooth conferencing
experience.
\end{itemize}
Enabling the above features for multi-party conferencing is challenging.
First, it requires a non-trivial formulation that allows systematic
solution design over arbitrary network capacity constraints. In contrast,
existing P2P system design works with performance guarantee commonly
assume bandwidth bottlenecks reside at the edge of the network. Second,
maximizing session rates subject to bounded delay is known to be NP-Complete
and hard to solve approximately \cite{vazirani2001approximation}.
We take a practical approach in this paper that explores all $2$-hop
delay-bounded overlay trees with polynomial complexity. Third, detecting
and reacting to network dynamics without \emph{a priori} knowledge
of the network conditions are non-trivial. We use both delay and loss
as congestion measures and adapt the session rates with respect to
both of them, allowing early detection and fast response to unpredictable
network dynamics.

The highlight in our design is a distributed rate control protocol,
that can discover and adapt to arbitrary topologies and network conditions
quickly, converging to efficient link rate allocations allowed by
the underlying network. In accordance with adaptive link rate control,
source video encoding rates are also dynamically controlled to optimize
video quality in arbitrary and unpredictable underlay network conditions.

The design of {\em Celerity} is largely inspired by our new formulation
that specifically takes into account arbitrary network capacity constraints
and allows us to explore design space beyond those in existing solutions.
Our formulation is overlay link based and has a number of variables
linear in the number of overlay links. This is a significant reduction
as compared to the number of variables exponential in the number of
overlay links in an alternative tree-based formulation. We believe
our approach is applicable to other P2P system problems, to allow
solution design beyond the common assumption in P2P scenarios that
the bandwidth bottlenecks reside only at the edge of the network.

We have implemented a prototype {\em Celerity} system using C++.
By extensive experiments in a local experimental testbed and on the
Internet, we demonstrate the superior performance of {\em Celerity}
over state-of-the-art solutions Simulcast and Mutualcast.

\subsection{Paper Organization}

The rest of this paper is organized as follows. In Section~\ref{sec:pf},
we introduce a general formulation for the multi-party conferencing
problem; existing solutions can be considered as algorithms solving
its special cases. We present and discuss the designs of two critical
components of {\em Celerity}, the tree packing module and the link
rate control module, in Sections \ref{sec:pt} and \ref{sec:rc},
respectively. We present the practical implementation of {\em Celerity}
in Section \ref{sec:impl} and the experimental results in Section
\ref{sec:expr}. Finally, we conclude in Section \ref{sec:conclusion}.
We leave all the proofs and pseudo codes in the Appendix.

\section{Problem Formulation and Celerity Overview}

\label{sec:pf}

One way to design a multi-party conferencing system is to formulate
its fundamental design problem, explore powerful theoretical techniques
to solve the problem, and use the obtained insights to guide practical
system designs. In this way, we can also be clear about potential
and limitation of the designs, allowing easy system tuning and further
systematic improvements. \rev{Table \ref{tab:symbol} lists the
key notations used in this paper.}
\begin{table}
\begin{tabular}{l|l}
\hline
\textbf{Notation}  & \textbf{Definition}\tabularnewline
\hline
$\mathcal{L}$  & Set of all physical links \tabularnewline
$V$  & Set of conference participating nodes \tabularnewline
$E$  & Set of directed overlay links \tabularnewline
$C_{l}$  & Capacity of the physical link $l$ \tabularnewline
$a_{l,e}$  & Whether overlay link $e$ passes physical link $l$\tabularnewline
$c_{m,e}$ & Rate allocated to session $m$ on overlay link $e$\tabularnewline
$\boldsymbol{c}_{m}$  & Overlay link rates of stream $m$, $\boldsymbol{c}_{m}=[c_{m,e},e\in E]$ \tabularnewline
$\boldsymbol{c}$  & Overlay link rates of all streams, $\boldsymbol{c}=[\boldsymbol{c}_{1}^{T},\ldots,\boldsymbol{c}_{M}^{T}]^{T}$\tabularnewline
$\boldsymbol{y}$  & Total overlay link traffics, $\boldsymbol{y}=\sum_{m=1}^{M}\boldsymbol{c}_{m}$ \tabularnewline
$D$  & Delay bound \tabularnewline
$R_{m}\left(\boldsymbol{c}_{m},D\right)$  & Session $m$'s rate within the delay bound $D$ \tabularnewline
$q_{l}(z)$  & Price function of violating link $l$'s capacity constraint\tabularnewline
$p_{l}$  & Lagrange multiplier of link $l$'s capacity constraint\tabularnewline
$\mathcal{G}\left(\boldsymbol{c},\boldsymbol{p}\right)$  & Lagrange function of variables $\boldsymbol{c}$ and $\boldsymbol{p}$\tabularnewline
\hline
\end{tabular}

Note: we use bold symbols to denote vectors%
. \caption{Key notations.}

\label{tab:symbol}
\end{table}

\subsection{Settings}

Consider a network modeled as a directed graph $G=\left(\mathcal{N},\mathcal{L}\right)$,
where {$\mathcal{N}$} is the set of all physical nodes, including
conference participating nodes and other intermediate nodes such as
routers, and $\mathcal{L}$ is the set of all physical links. Each
link $l\in\mathcal{L}$ has a nonnegative capacity $C_{l}$ and a
nonnegative propagation delay $d_{l}$.

Consider a multi-party conferencing system over $G$. We use $V\subseteq\mathcal{N}$
to denote the set of all conference participating nodes. Every node
in $V$ is a source and at the same time a receiver for every other
nodes. Thus there are totally $M\triangleq|V|$ sessions of (audio/video)
streams. Each stream is generated at a source node, say $v$, and
needs to be delivered to all the rest nodes in $V-\left\{ v\right\} $,
by using overlay links between any two nodes in $V$.

An overlay link $\left(u,v\right)$ means $u$ can send data to $v$
by setting up a TCP/UDP connection, along an underlay path from $u$
to $v$ pre-assigned by routing protocols. Let $E$ be the set of
all directed overlay links. For all $e\in E$ and $l\in\mathcal{L}$,
we define
\begin{equation}
a_{l,e}=\begin{cases}
1, & \mbox{if overlay link \ensuremath{e} passes physical link \ensuremath{l};}\\
0, & \mbox{otherwise.}
\end{cases}
\end{equation}
 The physical link capacity constraints are then expressed as
\[
\boldsymbol{a}_{l}^{T}\boldsymbol{y}=\sum_{e\in E}a_{l,e}\sum_{m=1}^{M}c_{m,e}\le C_{l},\quad\forall l\in\mathcal{L},
\]
 where $c_{m,e}$ denotes the rate allocated to session $m$ on overlay
link $e$ and $\boldsymbol{a}_{l}^{T}\boldsymbol{y}$ describes the
total overlay traffic passing through physical link $l$.

\textbf{Remark}: In our model, the capacity bottleneck can be anywhere
in the network, not necessarily at the edges. This is in contrast
to a common assumption made in previous P2P works that the uplinks/downlinks
of participating nodes are the only capacity bottleneck.

\subsection{Problem Formulation}

In a multi-party conferencing system, each session source broadcasts
its stream to all receivers over a complete overlay graph on which
every link $e$ has a rate $c_{m,e}$ and a delay $\sum_{l\in\mathcal{L}}a_{l,e}d_{l}$.
For smooth conferencing experience, the total delay of delivering
a packet from the source to any receiver, traversing one or multiple
overlay links, cannot exceed a delay bound $D$. %

A fundamental design problem is to maximize the overall conferencing
experience, by properly allocating the overlay link rates to the streams
subject to physical link capacity constraints. We formulate the problem
as a network utility maximization problem:
\begin{eqnarray}
\mathbf{MP}: & \max_{\boldsymbol{c}\ge0} & \sum_{m=1}^{M}U_{m}\left(R_{m}(\boldsymbol{c}_{m},D)\right)\label{eq:NUM1}\\
 & \text{s.t.} & \boldsymbol{a}_{l}^{T}\boldsymbol{y}\le C_{l},\quad\forall l\in\mathcal{L}.\label{eq:NUM2}
\end{eqnarray}
 The optimization variables are $\boldsymbol{c}$ and the constraints
in \eqref{eq:NUM2} are the physical link capacity constraints.

$R_{m}(\boldsymbol{c}_{m},D)$ denotes session $m$'s rate that we
obtain by using resource $\boldsymbol{c}_{m}$ \emph{within the delay
bound} $D$, and is a concave function of $\boldsymbol{c}_{m}$ as
we will show in Corollary~\ref{cor:Rm(cm)-is-concave} in the next
section.

The objective is to maximize the aggregate system utility. $U_{m}(R_{m})$
is an increasing and strictly concave function that maps the stream
rate to an application-specific utility. For example, a commonly used
video quality measure Peak Signal-to-Noise Ratio (PSNR) can be modeled
by using a logarithmic function as the utility \cite{chen2008ump}
\footnote{Using logarithmic functions also guarantees (weighted) proportional
fairness among sessions and thus no session will starve at the optimal
solution \cite{all:UtilityFunc:MW00}.%
}. With these settings and observations, $U_{m}(R_{m})$ is concave
in $\boldsymbol{c}$ and the problem \textbf{MP} is a concave optimization
problem.

\textbf{Remarks}: (i) The formulation of \textbf{MP} is an overlay
link based formulation in which the number of variables per session
is $|E|$ and thus at most $|V|^{2}$. One can write an equivalent
tree-based formulation for \textbf{MP} but the number of variables
per session will be \emph{exponential} in $|E|$ and $|V|$. (ii)
Existing solutions, such as Simulcast and Mutualcast, can be thought
as algorithms solving special cases of the problem \textbf{MP}. For
example, Simulcast can be thought as solving the problem \textbf{MP}
by using only the $1$-hop tree to broadcast content within a session.
Mutualcast can be thought as solving a special case of the problem
\textbf{MP} (with the uplinks of participating nodes being the only
capacity bottleneck) by packing certain depth-1 and depth-2 trees
within a session.

\subsection{Celerity Overview}

\label{subsec:co}

\emph{Celerity} builds upon two main modules to maximize the system
utility: (1) a {\em delay-bounded video delivery} module to distribute
video at high rate given overlay link rates (i.e., how to compute
and achieve $R_{m}(\boldsymbol{c}_{m},D)$); (2) a {\em link rate
control} module to determine $\boldsymbol{c}_{m}$.

\textbf{Video delivery under known link constraints}: This problem
is similar to the classic multicast problem, and packing spanning
(or Steiner) trees at the multicast source is a popular solution.
However, the unique {}``delay-bounded'' requirement in multi-party
conferencing makes the problem more challenging. We introduce a delay-bounded
tree packing algorithm to tackle this problem (detailed in Section~\ref{sec:pt}).

\textbf{Link rate control}: In principle, one can first infer the
network constraints and then solve the problem \textbf{MP} centrally.
However, directly inferring the constraints potentially requires knowing
the entire network topology and is highly challenging. In \emph{Celerity},
we resort to design adaptive and iterative algorithms for solving
the problem \textbf{MP} in a distributed manner, without \emph{a priori}
knowledge of the network conditions (detailed in Section~\ref{sec:rc}).

\begin{figure}[tb]
\centering \includegraphics[width=190pt]{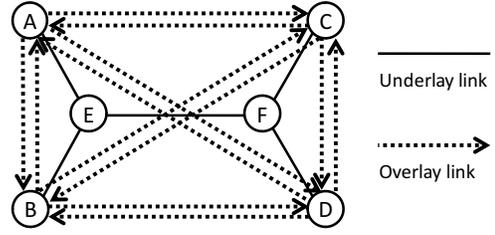}\\
 \caption{An illustrating example of 4-party ($A$, $B$, $C$, and $D$) conferencing
over a dumbbell underlay topology. $E$ and $F$ are two routers.
Solid lines represent underlay physical links. To make the graph easy
to read, we use one solid line to represent a pair of directed physical
links. Dash lines represent overlay links.}

\label{Fig:celerity.illustration}
\end{figure}

\rev{ We high-levelly explain how \emph{Celerity} works in a 4-party
conferencing example in Fig. \ref{Fig:celerity.illustration}. We
focus on session $A$, in which source $A$ distributes its stream
to receivers $B$, $C$, and $D$, by packing delay-bounded trees
over a complete overlay graph shown in the figure. We focus on source
$A$ and one overlay link $(B,C)$, which represents a UDP connection
over an underlay path $B$ to $E$ to $F$ to $C$. Other overlay
links and other sessions are similar.}

\rev{ We first describe the control plane operations. For the overlay
link $(B,C)$, the head node $B$ works with the tail node $C$ to
\emph{periodically} adjust the session rate $c_{A,B\rightarrow C}$
according to \emph{Celerity}'s link rate control algorithm. Such adjustment
utilizes control-plane information that source $A$ piggybacks with
data packets, and loss and delay statistics experienced by packets
traveling from $B$ to $C$. We show such local adjustments at every
overlay link result in globally optimal session rates. }

\rev{ The head node $B$ also \emph{periodically} reports to source
$A$ the session rate $c_{A,B\rightarrow C}$ and the end-to-end delay
from $B$ to $C$. Based on these reports from all overlay links,
source $A$ \emph{periodically} packs delay-bounded trees using \emph{Celerity}'s
tree-packing algorithm, calculates necessary control-plane information,
and delivers data and the control-plane information along the trees.}

\rev{ The data plane operations are simple. \emph{Celerity} uses
delay-bounded trees to distribute data in a session. Nodes on every
tree forward packets from its upstream parent to its downstream children,
following the {}``next-children'' tree-routing information embedded
in the packet header. \emph{Celerity}'s tree-packing algorithm guarantees
that (i) packets arrive at all receivers within the delay bound, and
(ii) the total rate of a session $m$ passing through an overlay link
$e$ does not exceed the allocated rate $c_{m,e}$. }

In the following two sections, we first present the designs of the
two main modules in \emph{Celerity}. We then describe how they are
implemented in physical peers in Section~\ref{sec:impl}.

\section{Packing Delay-bounded Trees}

\label{sec:pt}

Given the link rate vector $\boldsymbol{c}_{m}$ and delay for every
overlay link $e$ (i.e.,$\sum_{l\in\mathcal{L}}a_{l,e}d_{l}$), achieving
the maximum broadcast/multicast stream rate under a delay bound $D$
is a challenging problem. A general way to explore the broadcast/multicast
rate under delay bounds is to pack delay-bounded Steiner trees. However,
such problem is $NP$-hard \cite{guo2002qdmr}. Moreover, the number
of delay-bounded Steiner trees to consider is in general exponential
in the network size.

In this paper, we pack $2$-hop delay-bounded trees in an overlay
graph of session $m$, denoted by $\mathcal{D}_{m}$, to achieve a
good stream rate under a delay bound. Note by graph theory notations,
a $2$-hop tree has a depth at most $2$. Packing $2$-hop trees is
easy to implement. It also explores all overlay links between source
and receiver and between receivers, thus trying to utilize resource
efficiently. In fact, it is shown in \cite{chen2008ump,all:Mutualcast:LPZ05}
that packing $2$-hop multicast trees suffices to achieve the maximum
multicast rate for certain P2P topologies. We elaborate our tree-packing
scheme in the following.

\begin{figure}
\centering \includegraphics[width=4.8cm]{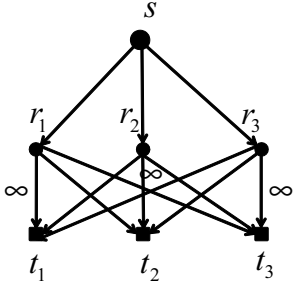}\\
 \caption{Illustration of the directed acyclic sub-graph over which we pack
delay-bounded $2$-hop trees.}

\label{fig:dag}
\end{figure}

We first define the overlay graph $\mathcal{D}_{m}$. Graph $\mathcal{D}_{m}$
is a directed acyclic graph with two layers; one example of such graph
is illustrated in Fig.~\ref{fig:dag}. In this example, consider
a session with a source $s$, three receivers $1,2,3$. For each receiver
$i$, we draw two nodes, $r_{i}$ and $t_{i}$, in the graph $\mathcal{D}_{m}$;
$t_{i}$ models the receiving functionality of node $i$ and $r_{i}$
models the relaying functionality of node $i$.

Suppose that the prescribed link bit rates are given by the vector
$\boldsymbol{c}_{m}$, with the capacity for link $e$ being $c_{m,e}$.
Then in $\mathcal{D}_{m}$, the link from $s$ to $r_{i}$ has capacity
$c_{m,s\rightarrow r_{i}}$, the link from $r_{i}$ to $t_{j}$ (with
$i\ne j$) has capacity $c_{m,r_{i}\rightarrow t_{j}}$, and the link
from $r_{i}$ to $t_{i}$ has infinite capacity. If the propagation
delay of an edge $e$ exceeds the delay bound, we do not include it
in the graph. If the propagation delay of a two-hop path $s\rightarrow r_{i}\rightarrow t_{j}$
exceeds the delay bound, we omit the edge from $r_{i}$ to $t_{j}$
from the graph. As a result, every path from $s$ to any receiver
$t_{i}$ in the graph has a path propagation delay within the delay
bound.


Over such $2$-layer sub-graph $\mathcal{D}_{m}$, we pack $2$-hop
trees connecting the source and every receiver using the greedy algorithm
proposed in~\cite{lovasz1976two}. Below we simply describe the algorithm
and more details can be found in~\cite{lovasz1976two}.

Assuming all edges have unit-capacity and allowing multiple edges
for each ordered node pair. The algorithm packs unit-capacity trees
one by one. Each unit-capacity tree is constructed by greedily constructing
a tree edge by edge starting from the source and augmenting towards
all receivers. It is similar to the greedy tree-packing algorithm
based on Prim's algorithm. The distinction lies in the rule of selecting
the edge among all potential edges. The edge whose removal leads to
least reduction in the multicast capacity of the residual graph is
chosen in the greedy algorithm.

\begin{figure*}[t!]
 \centering \includegraphics[width=18cm]{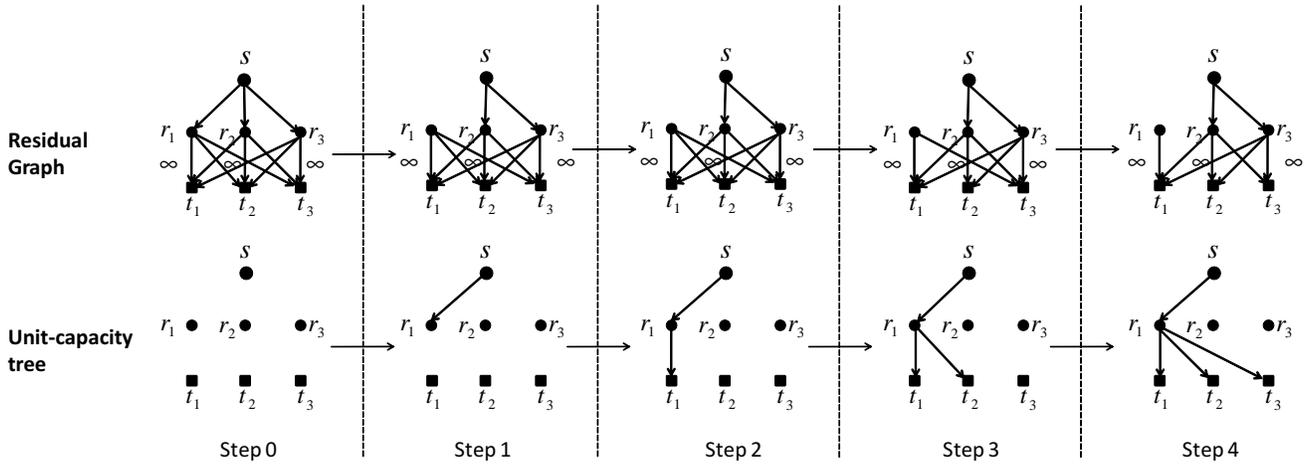} \caption{Example of packing a unit-capacity tree, starting from $s$ and reaching
all receivers $t_{1},t_{2}$ and $t_{3}$, using our greedy tree packing
algorithm. }

\label{Flo:example.tree.packing}
\end{figure*}

We show a simple example to illustrate how the tree packing algorithm
works. Fig. \ref{Flo:example.tree.packing} shows the process of packing
a unit-capacity tree over a $2$-layer sub-graph. In this example,
$s$ is source and $t_{1}$, $t_{2}$, $t_{3}$ are three receivers,
each edge from $s$ to $r_{i}\,(i=1,2,3)$ and from $r_{i}$ to $t_{j}\,(i\neq j)$
has unit capacity. The $\infty$ associated with the edge between
$r_{i}$ and $t_{i}$ means the edge has infinite capacity.

The tree packing algorithm maintains a {}``connected set'', denoted
by $\mathcal{T}$, that contains all the nodes that can be reached
from $s$ during the tree construction process. Initially, $\mathcal{T}=\{s\}$
contains only the source $s$. In each step, the algorithm adds and
connects one more node to the tree and appends the node into $\mathcal{T}$.
The algorithm finds a tree when $\mathcal{T}$ contains all the receivers.

Seen from Fig. \ref{Flo:example.tree.packing}, in Step 1, the algorithm
evaluates the links starting from source and greedily picks the edge
whose removal gives the smallest reduction of the multicast capacity
in the residual graph. In this example, any edge leaving $s$ can
be chosen because their removals give the same reduction. Our algorithm
randomly picks one such equally-good edge, in this case say edge $s\rightarrow r_{1}$.
The algorithm adds node $r_{1}$ into $\mathcal{T}$ and amends it
to be $\mathcal{T}=\{s,r_{1}\}$.

In Step 2, the algorithm evaluates the edges originated from any node
in $\mathcal{T}$. In this case it picks edge $r_{1}\rightarrow t_{1}$
and amends $\mathcal{T}$ to be $\{s,r_{1},t_{1}\}$. The algorithm
repeats the process until all the receivers are in $\mathcal{T}$,
which is Step 4 in this example. The algorithm then successfully constructs
a unit-capacity tree $s\rightarrow r_{1}\rightarrow\{t_{1},t_{2},t_{3}\}$.
Afterwards, the algorithm resets $\mathcal{T}=\{s\}$ and constructs
next tree in the residual graph until no unit-capacity tree can be
further constructed.

The above greedy algorithms is very simple to implement and its practical
implementation details are further discussed in Section~\ref{sec:impl}.

Utilizing the special structure of the graph $\mathcal{D}_{m}$, we
obtain performance guarantee of the algorithm as follows. \begin{thm}
The tree-packing algorithm in~\cite{lovasz1976two} achieves the
minimum of the min-cuts separating the source and receivers in $\mathcal{D}_{m}$
and is expressed as
\begin{align}
R_{m}(\boldsymbol{c}_{m},D)=\min_{j}\sum_{i}\min\left\{ c_{m,s\rightarrow r_{i}},c_{m,r_{i}\rightarrow t_{j}}\right\} .
\end{align}
 Furthermore, the algorithm has a running time of $O(|V||E|^{2})$.
\end{thm}

\textit{Proof:} Refer to Appendix A.

Hence, our tree-packing algorithm achieves the maximum delay-bounded
multicast rate over the $2$-layer sub-graph $\mathcal{D}_{m}$. The
achieved rate $R_{m}(\boldsymbol{c}_{m},D)$ is a concave function
of $\boldsymbol{c}_{m}$ as summarized below. \begin{cor} \label{cor:Rm(cm)-is-concave}The
delay-bounded multicast rate $R_{m}(\boldsymbol{c}_{m},D)$ obtained
by our tree-packing algorithm is a concave function of the overlay
link rates $\boldsymbol{c}_{m}$. \end{cor}

\textit{Proof:} Refer to Appendix B.

\com{should discuss total number of trees to consider is blabla
and compare the complexity of tree-based distributed solution and
link-based solution somewhere, perhaps in concluding remarks section.}

\subsection{Pack Delay-bounded Trees With Helpers Existing}

In the previous discussion, we do not involve helpers(a helper node
is neither a source nor a receiver in the conferencing session, but
it is willing to help in distributing content) in our tree packing
algorithm. Actually, this tree packing algorithm can also achieve
the minimum of the min-cuts separating the source and receivers in
$\mathcal{D}_{m}$ even though there exist helpers.

\begin{figure}
\centering \includegraphics[width=5.5cm]{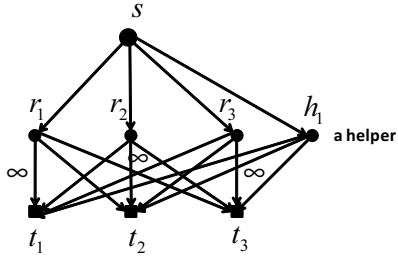}\\
 \caption{Illustration of the $2$-layer sub-graph $\mathcal{D}_{m}$ with a
helper existing}

\label{fig:dag-1}
\end{figure}

To see how the tree packing algorithm can be applied to $\mathcal{D}_{m}$
which includes helpers, we firstly define the 2-layer sub-graph $\mathcal{D}_{m}$
with helpers existing; one example of such graph is illustrated in
Fig.~\ref{fig:dag-1}. In this example, consider a session with a
source $s$, three receivers $1,2,3$, and a heper $h_{1}$. Similarly,
for each receiver $i$, we draw two nodes, $r_{i}$ and $t_{i}$,
in the graph $\mathcal{D}_{m}$; $t_{i}$ models the receiving functionality
of node $i$ and $r_{i}$ models the relaying functionality of node
$i$.

Suppose that the prescribed link bit rates are given by the vector
$\boldsymbol{c}_{m}$, with the capacity for link $e$ being $c_{m,e}$.
Then in $\mathcal{D}_{m}$, the link from $s$ to $r_{i}$ has capacity
$c_{m,s\rightarrow r_{i}}$, the link from $r_{i}$ to $t_{j}$ (with
$i\ne j$) has capacity $c_{m,r_{i}\rightarrow t_{j}}$, and the link
from $r_{i}$ to $t_{i}$ has infinite capacity. Similarly, the link
from $s$ to $h_{k}$(a helper) has capacity $c_{m,s\rightarrow h_{k}}$
and the link from $h_{k}$ to $t_{j}$ has capacity $c_{m,h_{k}\rightarrow t_{j}}$.
If the propagation delay of an edge $e$ exceeds the delay bound,
we do not include it in the graph. If the propagation delay of a two-hop
path $s\rightarrow v\,\left(v\in\left\{ r_{i}\right\} \cup\left\{ h_{k}\right\} \right)\rightarrow t_{j}$
exceeds the delay bound, we omit the edge from $v$ to $t_{j}$ from
the graph. As a result, every path from $s$ to any receiver $t_{j}$
in the graph has a path propagation delay within the delay bound.

Over such $2$-layer sub-graph $\mathcal{D}_{m}$, we use the same
greedy tree packing algorithm to pack $2$-hop trees connecting the
source and every receiver, and it can still achieve the minimum of
the min-cuts separating the source and receivers in $\mathcal{D}_{m}$
, which is discribed as follows.

\begin{thm} The tree-packing algorithm in~\cite{lovasz1976two}
achieves the minimum of the min-cuts separating the source and receivers
in $\mathcal{D}_{m}$ with helpers existing and is expressed as
\begin{align}
R_{m}(\boldsymbol{c}_{m},D)=\min_{j}\sum_{v\in\{r_{i}\}\cup\{h_{k}\}}\min\left\{ c_{m,s\rightarrow v},c_{m,v\rightarrow t_{j}}\right\} .
\end{align}
 Furthermore, the algorithm has a running time of $O(|V||E|^{2})$.
\end{thm}

\textit{Proof:} Refer to Appendix A.

Similarly, the achieved rate $R_{m}(\boldsymbol{c}_{m},D)$ is a concave
function of $\boldsymbol{c}_{m}$ as summarized below. \begin{cor}
\label{cor:Rm(cm)-is-concave-1}In the 2-layer sub-graph $\mathcal{D}_{m}$
with helpers existing, the delay-bounded multicast rate $R_{m}(\boldsymbol{c}_{m},D)$
obtained by our tree-packing algorithm is a concave function of the
overlay link rates $\boldsymbol{c}_{m}$. \end{cor}

\textit{Proof:} Refer to Appendix B.

\section{Overlay Link Rate Control}

\label{sec:rc}

\subsection{Considering Both Delay and Loss}

We revise original formulation to design our link rate control algorithm
with both queuing delay and loss rate taken into account. Adapting
link rates to both delay and loss allows early detection and fast
response to network dynamics.

Consider the following formulation with a penalty term added into
the objective function of the problem \textbf{MP}:
\begin{eqnarray}
\mathbf{MP-EQ}:\max_{\boldsymbol{c}\geq0} & \mathcal{U}(\boldsymbol{c})\stackrel{\Delta}{=} & \sum_{m=1}^{M}U_{m}\left(R_{m}(\boldsymbol{c}_{m},D)\right)-\sum_{l\in\mathcal{L}}\int_{0}^{\boldsymbol{a}_{l}^{T}\boldsymbol{y}}q_{l}(z)\, dz,\label{eq:primal.subgradient.dual1}\\
\text{s.t.} &  & \boldsymbol{a}_{l}^{T}\boldsymbol{y}\le C_{l},\quad\forall l\in\mathcal{L},\label{eq:primal.subgradient.dual2}
\end{eqnarray}
 where $\int_{0}^{\boldsymbol{a}_{l}^{T}\boldsymbol{y}}q_{l}(z)\, dz$
is the penalty associated with violating the capacity constraint of
physical link $l\in\mathcal{L}$, and we choose the price function
to be
\begin{align}
q_{l}(z) & \stackrel{\Delta}{=}\frac{(z-C_{l})^{+}}{z},\label{eq:P}
\end{align}
 where $(a)^{+}=\max\{a,0\}$. If all the constraints are satisfied,
then the second term in \eqref{eq:primal.subgradient.dual1} vanishes;
if instead some constraints are violated, then we charge some penalty
for doing so.

\textbf{Remark:} (i) The problem \textbf{MP-EQ} is equivalent to the
original problem \textbf{MP}. Because any feasible solution $\boldsymbol{c}$
of these two problems must satisfy $\boldsymbol{a}_{l}^{T}\boldsymbol{y}\leq C_{l}$,
and consequently the penalty term in the problem \textbf{MP-EQ} vanishes.
Therefore, any optimal solution of the original problem \textbf{MP}
must be an optimal solution of the problem \textbf{MP-EQ} and vice
versa. (ii) It can be verified that $-\sum_{l\in\mathcal{L}}\int_{0}^{\boldsymbol{a}_{l}^{T}\boldsymbol{y}}q_{l}(z)\, dz$
is a concave function in $\boldsymbol{c}$; hence, $\mathcal{U}(\boldsymbol{c})$
is a linear combination of concave functions and is concave. However,
because $R_{m}(\boldsymbol{c}_{m},D)$ is the minimum min-cut of the
overlay graph $\mathcal{D}_{m}$ with link rates being $\boldsymbol{c}_{m}$,
$\mathcal{U}(\boldsymbol{c})$ is not a differentiable function \cite{wu2006distributed}.

We apply Lagrange dual approach to design distributed algorithms for
the problem \textbf{MP-EQ}. The advantage of adopting distributed
rate control algorithms in our system is that it allows robust adaption
upon unpredictable network dynamics.

The Lagrange function of the problem is given by:
\begin{eqnarray}
\mathcal{G}\left(\boldsymbol{c},\boldsymbol{p}\right) & \triangleq & \sum_{m=1}^{M}U_{m}\left(R_{m}(\boldsymbol{c}_{m},D)\right)-\sum_{l\in\mathcal{L}}\int_{0}^{\boldsymbol{a}_{l}^{T}\boldsymbol{y}}q_{l}(z)\, dz-\nonumber \\
 &  & \sum_{l\in\mathcal{L}}p_{l}\left(\boldsymbol{a}_{l}^{T}\boldsymbol{y}-C_{l}\right),\label{eq:lag.function}
\end{eqnarray}
 where $p_{l}\geq0$ is the Lagrange multiplier associated with the
capacity constraint in (\ref{eq:primal.subgradient.dual2}) of physical
link $l$. $p_{l}$ can be interpreted as the price of using link
$l$. Since the problem \textbf{MP-EQ} is a concave optimization problem
with linear constraints, strong duality holds and there is no duality
gap. Any optimal solution of the problem and one of its corresponding
Lagrangian multiplier is a saddle point of $\mathcal{G}\left(\boldsymbol{c},\boldsymbol{p}\right)$
and vice versa. Thus to solve the problem \textbf{MP-EQ}, it suffices
to design algorithms to pursue saddle points of $\mathcal{G}\left(\boldsymbol{c},\boldsymbol{p}\right)$.

\subsection{A Loss-Delay Based Primal-Subgradient-Dual Algorithm}

There are two issues to address in designing algorithms for pursuing
saddle points of $\mathcal{G}\left(\boldsymbol{c},\boldsymbol{p}\right)$.
First, the utility function $\mathcal{U}(\boldsymbol{c})$ (and consequently
$\mathcal{G}\left(\boldsymbol{c},\boldsymbol{p}\right)$) is not everywhere
differentiable. Second, $\mathcal{U}(\boldsymbol{c})$ (and consequently
$\mathcal{G}\left(\boldsymbol{c},\boldsymbol{p}\right)$) is not strictly
concave in $\boldsymbol{c}$, thus distributed algorithms may not
converge to the desired saddle points under multi-party conferencing
settings \cite{chen2008ump}.

To address the first concern, we use subgradient in algorithm design.
To address the second concern, we provide a convergence result for
our designed algorithm.

To proceed, we first compute subgradients of $\mathcal{U}(\boldsymbol{c})$.
The proposition below presents a useful observation. \begin{prop}
A subgradient of $\mathcal{U}(\boldsymbol{c})$ with respect to $c_{m,e}$
for any $e\in E$ and $m=1,\ldots M$ is given by
\[
U'_{m}\left(R_{m}\right)\frac{\partial R_{m}}{\partial c_{m,e}}-\sum_{l\in\mathcal{L}}a_{l,e}\frac{(\boldsymbol{a}_{l}^{T}\boldsymbol{y}-C_{l})^{+}}{\boldsymbol{a}_{l}^{T}\boldsymbol{y}}
\]
 where $\frac{\partial R_{m}}{\partial c_{m,e}}$ is a subgradient
of $R_{m}(\boldsymbol{c}_{m},D)$ with respect to $c_{m,e}$. \end{prop}
\textit{Proof:} Refer to Appendix C.

Motivated by the pioneering work of Arrow, Hurwicz, and Uzawa \cite{arrow1958studies}
and the followup works \cite{nedic2009subgradient}\cite{bruck1977weak},
we propose to use the following \emph{primal-subgradient-dual} algorithm
to pursue the saddle point of $\mathcal{G}\left(\boldsymbol{c},\boldsymbol{p}\right)$:$\forall e\epsilon E,\, m=1,...M,\,\forall l\epsilon\mathcal{L}$,\\
 \textbf{Primal-Subgradient-Dual Link Rate Control Algorithm:}
\begin{eqnarray}
c_{m,e}^{(k+1)} & = & c_{m,e}^{(k)}+\alpha\left[U'_{m}\left(R_{m}^{(k)}\right)\frac{\partial R_{m}^{(k)}}{\partial c_{m,e}}\right.\nonumber \\
 &  & \left.\sum_{l\in\mathcal{L}}a_{l,e}\frac{(\boldsymbol{a}_{l}^{T}\boldsymbol{y}^{(k)}-C_{l})^{+}}{\boldsymbol{a}_{l}^{T}\boldsymbol{y}^{(k)}}-\sum_{l\in\mathcal{L}}a_{l,e}p_{l}^{(k)}\right]_{c_{m,e}^{(k)}}^{+}\label{eq:subgradient.primal.dual.control1}\\
p_{l}^{(k+1)} & = & p_{l}^{(k)}+\frac{1}{C_{l}}\left[\boldsymbol{a}_{l}^{T}\boldsymbol{y}^{(k)}-C_{l}\right]_{p_{l}^{(k)}}^{+}\label{eq:subgradient.primal.dual.control2}
\end{eqnarray}
 where $\alpha>0$ represents a constant the step size for all the
iterations, and function
\[
[b]_{a}^{+}=\begin{cases}
\max(0,b), & a\le0;\\
b, & a>0.
\end{cases}
\]

We have the following observations for the control algorithm in (\ref{eq:subgradient.primal.dual.control1})-(\ref{eq:subgradient.primal.dual.control2}):
\begin{itemize}
\item It is known that $\sum_{l\in\mathcal{L}}a_{l,e}\frac{(\boldsymbol{a}_{l}^{T}\boldsymbol{y}-C_{l})^{+}}{\boldsymbol{a}_{l}^{T}\boldsymbol{y}}$
can be interpreted as the packet loss rate observed at overlay link
$e$ \cite{kelly2003fairness}. The intuitive explanation is as follows.
The term $(\boldsymbol{a}_{l}^{T}\boldsymbol{y}-C_{l})^{+}$ is the
excess traffic rate offered to physical link $l$; thus $\frac{(\boldsymbol{a}_{l}^{T}\boldsymbol{y}-C_{l})^{+}}{\boldsymbol{a}_{l}^{T}\boldsymbol{y}}$
models the fraction of traffic that is dropped at $l$. Assuming the
packet loss rates are additive (which is a reasonable assumption for
low packet loss rates), the total packet loss rates seen by the overlay
link $e$ is given by $\sum_{l\in\mathcal{L}}a_{l,e}\frac{(\boldsymbol{a}_{l}^{T}\boldsymbol{y}-C_{l})^{+}}{\boldsymbol{a}_{l}^{T}\boldsymbol{y}}$.
\item It is also known that $p_{l}$ updating according to (\ref{eq:subgradient.primal.dual.control2})
can be interpreted as queuing delay at physical link $l$ \cite{all:Vegas:LPW02}.
Intuitively, if the incoming rate $\boldsymbol{a}_{l}^{T}\boldsymbol{y}>C_{l}$
at $l$, then it introduces an additional queuing delay of $\frac{\boldsymbol{a}_{l}^{T}\boldsymbol{y}-C_{l}}{C_{l}}$
for $l$. If otherwise the term $\boldsymbol{a}_{l}^{T}\boldsymbol{y}\leq C_{l}$,
then the present queueing delay is reduced by an amount of $\frac{C_{l}-\boldsymbol{a}_{l}^{T}\boldsymbol{y}}{C_{l}}$
unless hitting zero. The total queuing delay observed by the overlay
link $e$ is then given by the sum $\sum_{l\in\mathcal{L}}a_{l,e}p_{l}$.
\item It turns out that the utility function, the subgradients, packet loss
rate and queuing delay are sufficient statistics to update $c_{m,e}$
independently of the updates of other link rates. This way, we can
solve the problem \textbf{MP-EQ} without knowing the physical network
topology and physical link capacities.
\end{itemize}
The algorithm in (\ref{eq:subgradient.primal.dual.control1})-(\ref{eq:subgradient.primal.dual.control2})
is similar to the standard primal-dual algorithm, but since $\mathcal{U}(\boldsymbol{c})$
is not differentiable everywhere, we use subgradient instead of gradient
in updating the overlay link rates $\boldsymbol{c}$. If we fix the
dual variables $\boldsymbol{p}$, then the algorithm in (\ref{eq:subgradient.primal.dual.control1})
corresponds to the standard subgradient method \cite{bertsekas1999nonlinear}.
It maximizes a non-differentiable function in a way similar to gradient
methods for differentiable functions --- in each step, the variables
are updated in the direction of a subgradient. However, such a direction
may not be an ascent direction; instead, the subgradient method relies
on a different property. If the variable takes a sufficiently small
step along the direction of a subgradient, then the new point is closer
to the set of optimal solutions.

Establishing convergence of subgradient algorithms for saddle-point
optimization is in general challenging \cite{nedic2009subgradient}.
We explore convergence properties for our primal-subgradient-dual
algorithm in the following theorem. %

\begin{thm} Let $(\boldsymbol{c}^{*},\boldsymbol{p}^{*})$ be a saddle
point of $\mathcal{G}\left(\boldsymbol{c},\boldsymbol{p}\right)$,
and $\bar{\mathcal{G}}^{(k)}$ be the average function value obtained
by the algorithm in (\ref{eq:subgradient.primal.dual.control1})-(\ref{eq:subgradient.primal.dual.control2})
after $k$ iterations:
\[
\bar{\mathcal{G}}^{(k)}\triangleq\frac{1}{k}\sum_{i=0}^{k-1}\mathcal{G}\left(\boldsymbol{c}^{(k)},\boldsymbol{p}^{(k)}\right).
\]
 Suppose $\left|U_{m}^{'}(R_{m}(\boldsymbol{c}_{m}))\right|\leq\bar{U}$,
$\forall m=1,\ldots,M$, where $\bar{U}$ is a constant, then we have
\begin{eqnarray*}
-\frac{B_{1}}{2\alpha k}-\frac{\Delta^{2}}{2}\alpha\leq & \bar{\mathcal{G}}^{(k)}-\mathcal{G}\left(\boldsymbol{c}^{*},\boldsymbol{p}^{*}\right) & \leq\frac{B_{2}}{2k}+\frac{\Delta^{2}}{2}\max_{l\in\mathcal{L}}C_{l}^{-1},
\end{eqnarray*}
 where $B_{1}=\left\Vert \boldsymbol{c}^{(0)}-\boldsymbol{c}^{*}\right\Vert ^{2}$
and $B_{2}=\left[\boldsymbol{p}^{(0)}-\boldsymbol{p}^{*}\right]^{T}\mbox{diag}\left(C_{l},l\in\mathcal{L}\right)\left[\boldsymbol{p}^{(0)}-\boldsymbol{p}^{*}\right]$
are two positive distances depending on $(\boldsymbol{c}^{(0)},\boldsymbol{p}^{(0)})$,
and $\Delta$ is a positive constant depending on $\bar{U}$ and $(\boldsymbol{c}^{(0)},\boldsymbol{p}^{(0)})$.
\end{thm} \textit{Proof:} Refer to Appendix D.

\textbf{Remarks}: (i) The results bound the time-average Lagrange
function value obtained by the algorithm to the optimal in terms of
distances of the initial iterates $(\boldsymbol{c}^{(0)},\boldsymbol{p}^{(0)})$
to a saddle point. In particular, the averaged function values $\bar{\mathcal{G}}^{(k)}$
converge to the saddle point value $\mathcal{G}\left(\boldsymbol{c}^{*},\boldsymbol{p}^{*}\right)$
within a gap of $\max\left(\alpha,\max_{l\in\mathcal{L}}C_{l}^{-1}\right)\frac{\Delta^{2}}{2}$,
at a rate of $1/k$. (ii) The requirement of the utility function
is easy to satisfied; one example is $U_{m}(z)=\log(z+\epsilon)$
with $\epsilon>0$. (iii) Our results generalize the one in \cite{nedic2009subgradient}
in the sense that the one in \cite{nedic2009subgradient} only applies
to the case of uniform step size, while we allow different $p_{l}$
to update with different step size $\frac{1}{C_{l}}$, which is critical
for $p_{l}$ to be interpreted as queuing delay and thus practically
measurable. Our results also have less stringent requirement on the
utility function than the one in \cite{nedic2009subgradient}. (iv)
Although the results may not warranty convergence in the strict sense,
our experiments over LAN testbed and on the Internet in Section~\ref{sec:expr}
show the algorithm quickly stabilizes around optimal operating points.
Obtaining stronger convergence results that confirm our practical
observations are of great interests and is left for future work.


\subsection{Computing Subgradients of $R_{m}(\boldsymbol{c}_{m},D)$}

\label{ssec:subgradient.calculation}

A key to implementing the Primal-Subgradient-Dual algorithm is to
obtain subgradients of $R_{m}(\boldsymbol{c}_{m},D)$. We first present
some preliminaries on subgradients, as well as concepts for computing
subgradients for $R_{m}(\boldsymbol{c}_{m},D)$.

\begin{defn} Given a convex function $f$, a vector $\xi$ is said
to be a subgradient of $f$ at $x\in\mathbf{dom}f$ if
\[
f(x')\geq f(x)+\xi^{T}(x'-x),\forall x'\in\mathbf{dom}f,
\]
 where $\mathbf{dom}f=\left\{ x\in\mathbf{R}^{n}||f(x)|<\infty\right\} $
represents the domain of the function $f$.

\end{defn}

For a concave function $f$, $-f$ is a convex function. A vector
$\xi$ is said to be a subgradient of $f$ at $x$ if $-\xi$ is a
subgradient of $-f$.

Next, we define the notion of a {\em critical cut}. For session
$m$, let its source be $s_{m}$ and receiver set be $V_{m}\subset V-\left\{ s_{m}\right\} $.
A partition of the vertex set, $V=Z\cup\bar{Z}$ with $s_{m}\in Z$
and $t\in\bar{Z}$ for some $t\in V_{m}$, determines an $s_{m}$-$t$-cut.
Define
\[
\delta(Z)\triangleq\left\{ \left(i,j\right)\in E|i\in Z,j\in\bar{Z}\right\}
\]
 be the set of overlay links originating from nodes in set $Z$ and
going into nodes in set $\bar{Z}$. Define the capacity of cut $(Z,\bar{Z})$
as the sum capacity of the links in $\delta(Z)$:
\[
\rho(Z)\triangleq\sum_{e\in\delta(Z)}c_{m,e}.
\]
 \begin{defn} For session $m$, a cut $(Z,\bar{Z})$ is an $s_{m}$-$V_{m}$
critical cut if it separates $s_{m}$ and any of its receivers and
$\rho(Z)=R_{m}(\boldsymbol{c}_{m},D)$. \end{defn}
\begin{figure}[h]

\begin{centering}
\includegraphics[width=3.8cm]{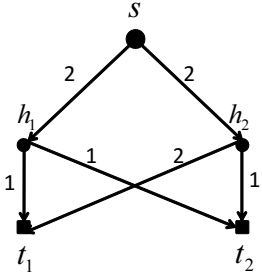}
\par\end{centering}

\caption{Critical cut example. Source $s$ and its two receivers $t_{1},t_{2}$
are connected over a directed graph. The number associated with a
link represents its link capacity.}

\label{Fig:example.critical cut}
\end{figure}

We show an example to illustrate the concept of critical cut. In Fig.
\ref{Fig:example.critical cut}, $s$ is a source, and $t_{1}$, $t_{2}$
are its two receivers. The minimum of the min-cuts among the receivers
is $2$. For the cut $(\{s,h_{1},h_{2},t_{1}\},\{t_{2}\})$, its $\delta(\{s,h_{1},h_{2},t_{1}\})$
contains links $(h_{1},t_{2})$ and $(h_{2},t_{2})$, each having
capacity one. Thus the cut $(\{s,h_{1},h_{2},t_{1}\},\{t_{2}\})$
has a capacity of $2$ and it is an $s-(t_{1},t_{2})$ critical cut.

With necessary preliminaries, we turn to compute subgradients of $R_{m}(\boldsymbol{c}_{m},D)$.
Since $R_{m}(\boldsymbol{c}_{m},D)$ is the minimum min-cut of $s_{m}$
and its receivers over the overlay graph $\mathcal{D}_{m}$, it is
known that one of its subgradients can be computed in the following
way \cite{wu2006distributed}.
\begin{itemize}
\item Find an $s_{m}$-$V_{m}$ critical cut for session $m$, denote it
as $(Z,\bar{Z})$. Note there can be multiple $s_{m}$-$V_{m}$ critical
cuts in graph $\mathcal{D}_{m}$, and it is sufficient to find any
one of them. 

\item A subgradient of $R_{m}(\boldsymbol{c}_{m},D)$ with respect to $c_{m,e}$
is given by
\begin{equation}
\frac{\partial R_{m}(\boldsymbol{c}_{m},D)}{\partial c_{m,e}}=\begin{cases}
1, & \mbox{if \ensuremath{e\in\delta(Z)};}\\
0, & \mbox{otherwise.}
\end{cases}\label{eq:subgradient}
\end{equation}

\end{itemize}
In our system, these subgradients are computed by the source of each
session, after collecting the overlay-link rates from each receiver
in the session. More implementation details are in Section~\ref{sec:impl}.

\section{PRACTICAL IMPLEMENTATION}

\label{sec:impl}

Using the asynchronous networking paradigm supported by the asynchronous
I/O library (called \texttt{asio}) in the \texttt{Boost} C++ library,
we have implemented a prototype of \emph{Celerity}, our proposed multi-party
conferencing system, with about $17,000$ lines of code in C++.

\begin{figure}[H]
 \centering\includegraphics[width=7.5cm]{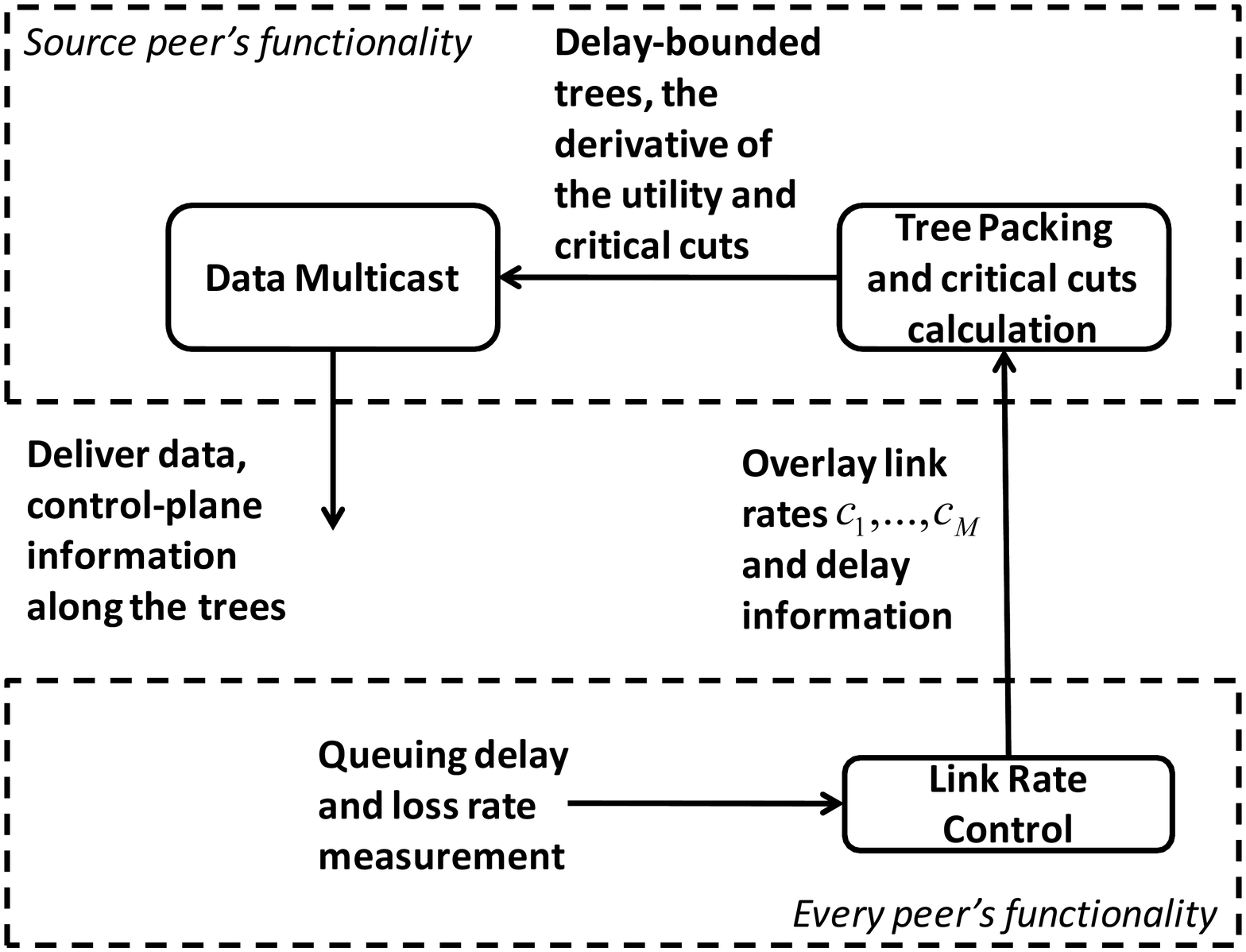}

\caption{System architecture of \emph{Celerity}.}

\label{Flo:sys.overview}
\end{figure}

\emph{Celerity} consists of three main modules: link rate
control module, tree-packing and critical cut calculation module,
and the data multicast engine. Fig. \ref{Flo:sys.overview} describes
the relationship between these components and where they physically
reside.

In the following, we describe the functionality implemented by peers,
some critical implementations, operation overhead and the peer computation overhead.

\subsection{Peer Functionality}
\label{PeerFun}

In our implementation, all peers perform the
following functions:
\begin{itemize}
\item Peers in broadcast trees forward packets received from its upstream
parent to its downstream children. Sufficient information about downstream
children in the tree is embedded in the packet header, for a packet
to become {}``self-routing\textquotedblright{} from the source to
all leaf nodes in a tree.
\item Every 200 ms, each peer calculates the loss rate and queuing delay
of its incoming links and adjusts the rates of its incoming links
based on the link rate control algorithm, and then sends them to their
corresponding upstream senders for the new rates to take effect.
\item Every 300 ms, each peer sends the link state (including allocated
rate and Round Trip Time) of all its outgoing links for each session
to the source of the session.
\end{itemize}
Upon receiving link states for all the links, the source of each session
uses the received link rates and the delay information to pack a new
set of delay-bounded trees, and starts transmitting session packets
along these trees. We set the delay bound to be $200$ ms when packing
delay-bounded trees in our implementation. When a source packs delay-bounded
trees, it also calculates one critical cut and the derivative of the
utility for its session based on the allocated link rates and the
delay information. In addition, the source embeds the information
about the critical cut and the derivative of the utility in the header
of outgoing packets. When these packets are received, a peer learns
the derivative of the utility and whether a link belongs to the critical
cut or not; it then adjusts the link rate accordingly.

In the following,
We use the example in Fig. \ref{Fig:celerity.illustration} to further
explain how \textit{Celerity} works.

For an overlay link $e\in E$, say $B\rightarrow C$, The tail node
$C$ is responsible for controlling $c_{A,e}$, the rate allocated
to session $A$. To do so, $C$ works with the head node $B$ to measure
the packet loss rate and queuing delay experienced by session $A$'s
packets over $e$ ($B\rightarrow C$). This can be done by $B$ attaching
local sequence numbers and timestamps to session $A$'s packets and
$C$ calculating the missing sequence numbers and the one-way-delay
based on the timestamps~\cite{chen2008ump}. $C$ also receives other
needed control plane information from the source of session $A$,
such as the critical cut information and the derivative of the utility,
along with the data packets arrived at $C$. With the loss rate and
queuing delay for session $A$'s packets, as well as these control
plane information, $C$ adjusts the allocated rate $c_{A,B\rightarrow C}$
using the algorithm in \eqref{eq:subgradient.primal.dual.control1}-\eqref{eq:subgradient.primal.dual.control2}
and sends it to $B$ for the new rate to take effect.

Every 300ms, The head node of each overlay link $e$ reports the allocated
rates $c_{m,e}$ and the overlay link round-trip-time information
to the source peers. Take the overlay link $B\rightarrow C$ for example,
$B$ reports the allocated rate $c_{A,B\rightarrow C}$ and the round-trip-time
information of this link to source $A$. With the collected link state
information, source peer $A$ packs delay-bounded trees using the
algorithm described in Section~\ref{sec:pt}, calculates critical
cuts using the method explained in Section~\ref{ssec:subgradient.calculation}
and the derivative of the utility, and then delivers data and the
control-plane information to the peers along the trees.

\subsection{Critical Cut Calculation}

The calculation of critical cuts, i.e., the subgradient of $R_{m}(\boldsymbol{c}_{m},D)$,
is the key to our implementation of the primal subgradient algorithm.
There can be multiple critical cuts in one session, but it is sufficient
to find any one of them. Since the source collects allocated rates
of all overlay links in its own session, it can calculate the min-cut
from the source to every receiver, and record the cut that achieves
the min-cut. Then, the source compares the capacities of these min-cuts,
and the cut with the smallest capacity is a critical cut.

\subsection{Utility Function}

With respect to the utility function in our prototype implementation,
the PSNR (peak signal-to-noise ratio) metric is the de facto standard
criterion to provide objective quality evaluation in video processing.
We observed that the PSNR of a video stream coded at a rate $z$ can
be approximated by a logarithmic function $\beta\log(z+\delta)$,
in which a higher $\beta$ represents videos with a larger amount
of motion. $\delta$ is a small positive constant to ensure the function
has a bounded derivative for $z\geq0$. Due to this observation, we
use a logarithmic utility function in our implementation.

\subsection{Opportunistic Local Loss Recovery}

Providing effective loss recovery in a delay-bounded reliable broadcast
scenario, such as multi-party conferencing, is known to be challenging
\cite{park2006codecast}. It is hard for error control coding to work
efficiently, since different receivers in a session may experience
different loss rates and thus choosing proper error control coding
parameters to avoid unnecessary waste of throughput is non-trivial.
If re-broadcasting the lost-packets is in use, it introduces additional
delay and may cause packets missing deadlines and become useless.

In our implementation, we use network coding \cite{park2006codecast}\cite{ahlswede2000network}
to allow flexible and opportunistic local loss recovery. For each
overlay link $e$, if the trees of a session $m$ do not exhaust $c_{m,e}$,
the overlay-link rate dedicated for the session, then we send coded
packets (i.e., linear combination of received packets of corresponding
session) over such link $e$. As such, receiver of the overlay link
$e$ can recover the packets that are lost on link $e$ locally by
using the network coded packets. This way, Celerity provides certain
flexible local loss recovery capability without incurring delay due
to retransmission.

\subsection{Fast Bootstrapping}
\label{ssec:fastBootstrapping}

Similar to TCP's Slow Start strategy, we implement a method in \emph{Celerity}
called {}``quick start\textquotedblright{} to quickly ramp up the
rates of all sessions during conference initialization stage. The
purpose is to quickly bootstrap the system to close-to-optimal operating
points when the conference just starts, during which period peers
are joining the conference and nothing significant is going on. We
achieve this by using larger values for $\beta$ in the utility functions
and a large step size in link rate adaptation during the first 30
seconds. After the initialization stage, we reset $\beta$ and step
sizes to proper values and allow our system converge gradually and
avoid unnecessary performance fluctuation.

\subsection{Operation Overhead}

There are two types of overhead in \textit{Celerity}: (1) \textit{packet
overhead}: the size of the application-layer packet header is around
$46$ bytes per data packet, including critical cut information, the
derivative of the utility, packet sequence number, coding vector,
timestamp and so on. (2) \textit{link-rate control and link-state
report overhead}: every $200$ ms, each peer adjusts the rates of
its incoming links and sends them to their corresponding upstream
senders. In our implementation, such rate-control overhead is $0.2$
kbps per link per session. For the link state report overhead, each
peer sends the link state of all its outgoing links for each session
to the source of the session every $300$ ms. In our implementation,
for each peer, such link-state report overhead is $0.158$ kbps per
link per session. In Section \ref{ssec:expr:internet}, we report
an overall operational overhead of $3.9$\% in our 4-party Internet
experiment.

\subsection{Peer Computation Overhead}
As described in Section \ref{PeerFun}, each peer in \emph{Celerity} delivers its own packets, forwards packets from other sessions, calculates the loss and queuing delay, updates the link rate of its incoming links, and reports the link states. In the worst case, a peer delivers its packets and forwards packets from other sessions to other peers using Simulcast. Thus for each peer the computation overhead of delivering and forwarding packets is $O(R|V||E|)$ per second, where $R$ is the maximum of $R_{m}(\boldsymbol{c}_{m},D)$ of all the sessions. For calculating the loss and queuing delay, each peer calculates the loss and queuing delay of its incoming links every 200 ms. Since the conferencing participants are fully connected by the overlay links, the computation overhead of this action is $O(|V|)$ per second per peer. Each peer updates the allocated link rate for each session of its incoming links and sends them to its upstreams every 200 ms. Since each incoming link is shared by all sessions, for each incoming link the peer should send $|V|$ link rate updating packets to the corresponding upstream. Thus the computation overhead of updating link rate is $O(|V||E|)$ per second per peer. Every 300 ms, each peer sends the link states of all its outgoing links for each session to the corresponding session source. Similarly, because each outgoing link is shared by all sessions and all the peers are fully connected, the computation overhead of reporting the links states is $O(|V||E|)$ per second for each peer. In addition, each peer packs trees and calculates critical cuts every 300 ms, according to Theorem 1, the computation complexity of these two actions are $O(|V||E|^{2})$ and $O(|V||E|)$ respectively. Thus the computation overhead of packing trees and calculating critical cuts is $O(|V||E|^{2})$ per second per peer. By summing up all these computation overheads, the overall computation overhead of each peer is $O(|V||E|^{2}+R|V||E|)$ per second.

\section{Experiments}

\label{sec:expr}

We evaluate our prototype \emph{Celerity} system over a LAN testbed
as well as over the Internet. The LAN experiments allow us to (i)
stress-test \emph{Celerity} under various network conditions; (ii)
see whether \emph{Celerity} meets the design goal -- delivering high
delay-bounded throughput and automatically adapting to dynamics in
the network; (iii) demonstrate the fundamental performance gains over
existing solutions, thus justifying our theory-inspired design.

The Internet experiments allow us to further access \emph{Celerity}'s
superior performance over existing solutions in the real world.

\subsection{LAN Testbed Experiments}

\begin{figure}[hbt]
 \centering{}\includegraphics[width=5.5cm]{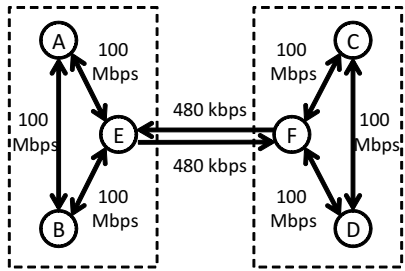}\caption{The {}``tough'' dumbbell topology of the experimental testbed. Two
conference participating nodes $A$ and $B$ are in one {}``office''
and another twos nodes $C$ and $D$ are in a different {}``office''.
The two {}``offices'' are connected by directed links between gateway
nodes $E$ and $F$, each link having a capacity of $480$ kbps. Link
propagation delays are negligible. }

\label{Fig:topology}
\end{figure}

We evaluate \emph{Celerity} over a LAN testbed illustrated in Fig.
\ref{Fig:topology}, where four PC nodes ($A,B,C,D$) are connected
over a LAN dumbbell topology. The dumbbell topology represents a popular
scenario of multi-party conferencing between branch offices. It is
also a {}``tough'' topology -- existing approaches, such as Simulcast
and Mutualcast, fail to efficiently utilize the bottleneck bandwidth
and optimize system performance.

In our experiments, all four nodes run \emph{Celerity}. We run a four-party
conference for 1000 seconds and evaluate the system performance. In
order to evaluate \emph{Celerity}'s performance in the presence of
network dynamics, we reduce cross traffic and introduce link failures
during the experiment. In particular, we introduce an 80kpbs cross-traffic
from node $E$ to node $F$ between the 300th second and the 500th
second, reducing the available bandwidth between $E$ and $F$ from
480 kbps to 400 kbps. Further, starting from the 700th second, we
disconnect the physical link between $A$ and $E$; this corresponds
to a practical situation where node $A$ suddenly cannot directly
communicate with nodes outside the {}``office'' due to middleware
or configuration errors at the gateway $E$.

\begin{figure*}[ht!]
\subfloat[Rate Performance of Node A]{\includegraphics[width=6cm]{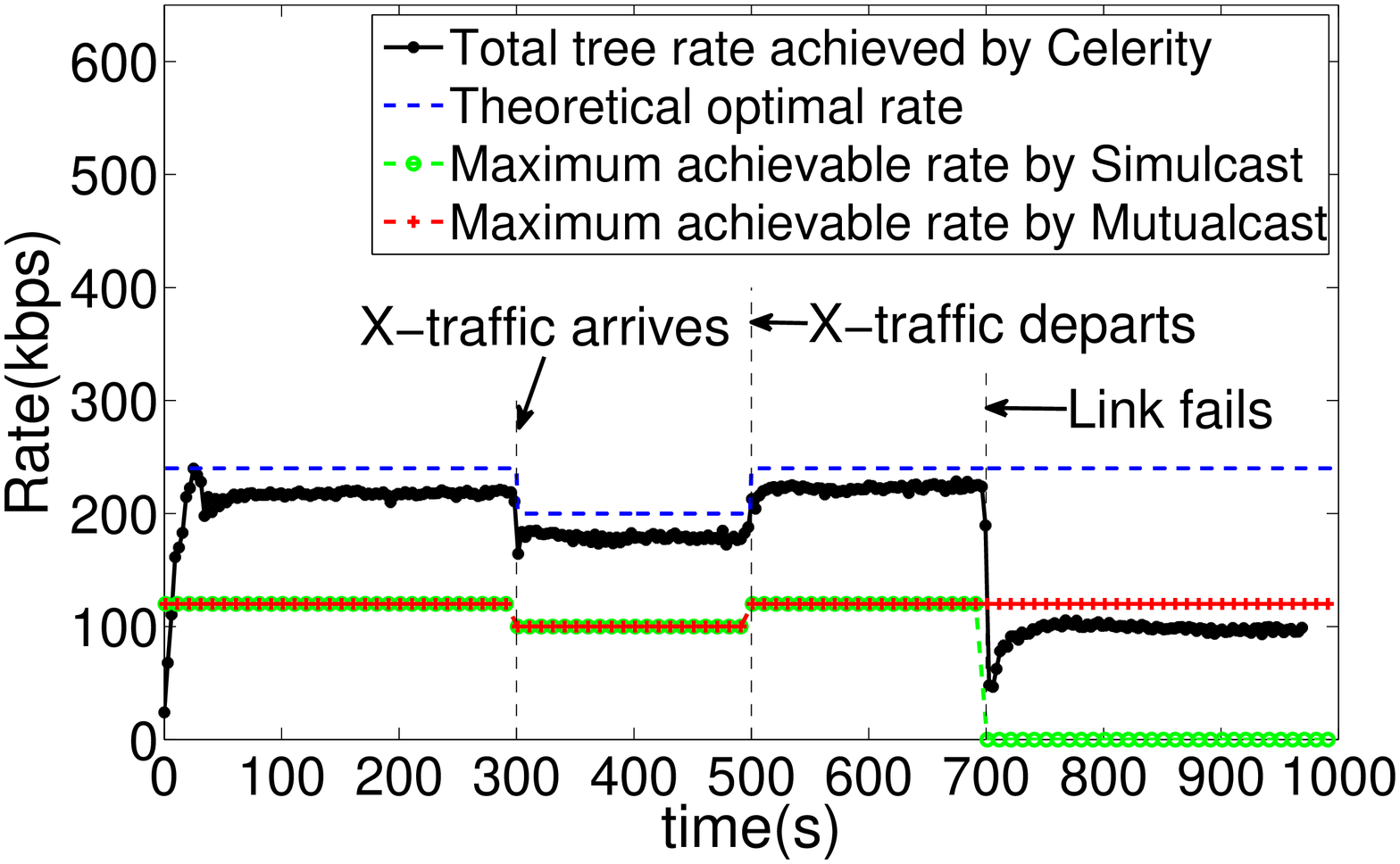}\label{Fig:LAN_rate_A}}\subfloat[Rate Performance of Node B]{\includegraphics[width=6cm]{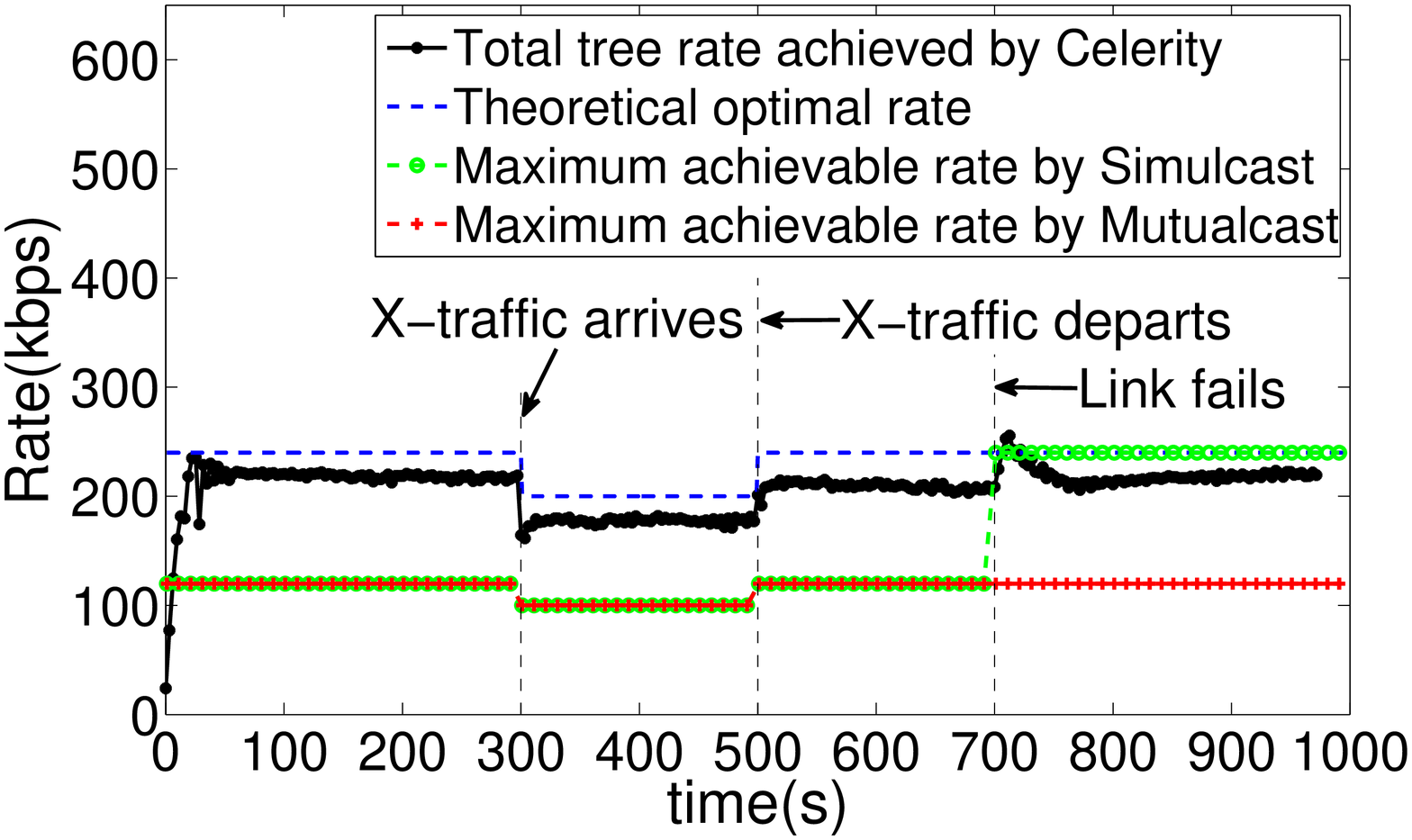}\label{Fig:LAN_rate_B}}\subfloat[Rate Performance of Node C]{\includegraphics[width=6cm]{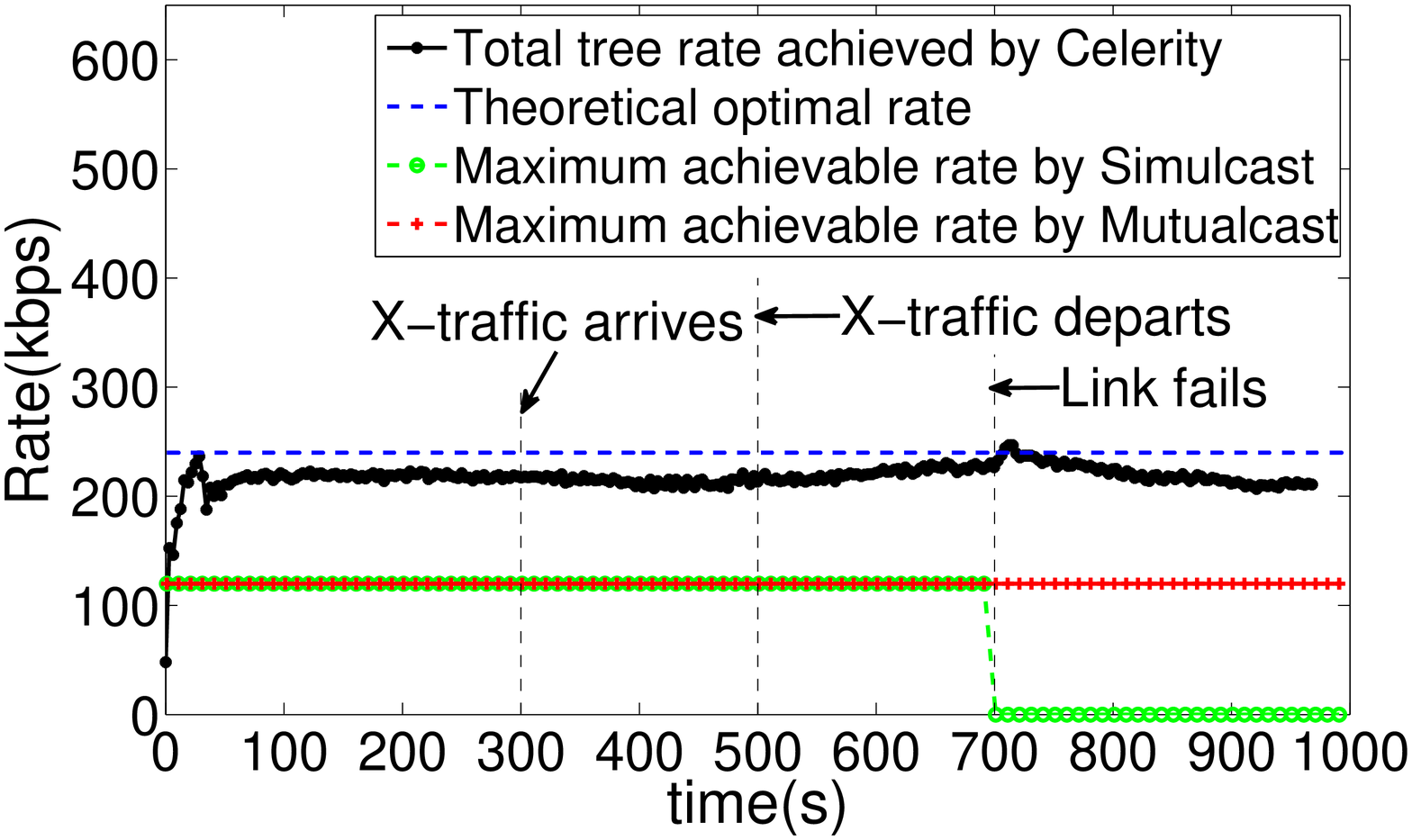}\label{Fig:LAN_rate_C}}

\subfloat[Rate Performance of Node D]{\includegraphics[width=6cm]{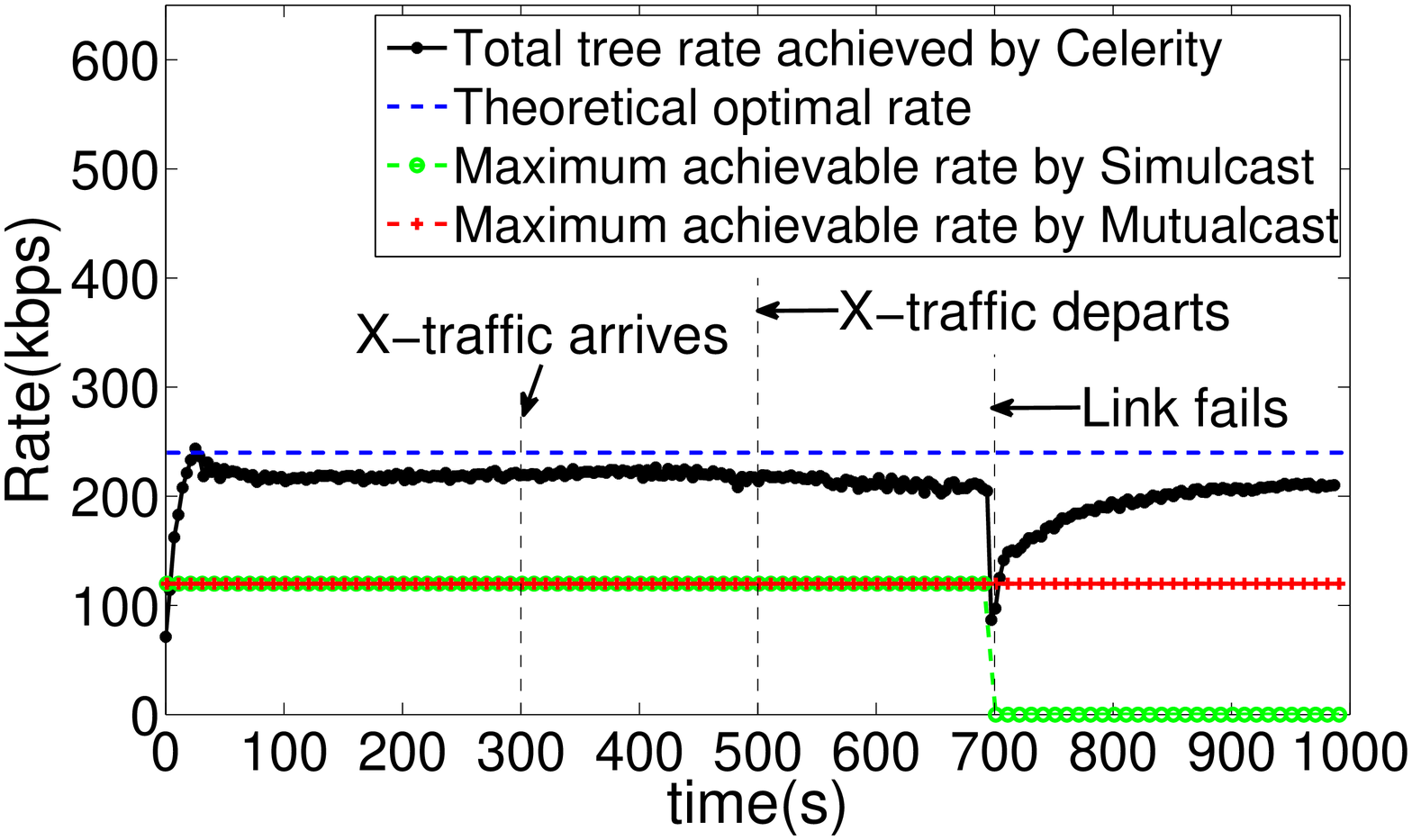}

\label{Fig:LAN_rate_D}}\subfloat[Total utility of all sessions]{\includegraphics[width=6cm]{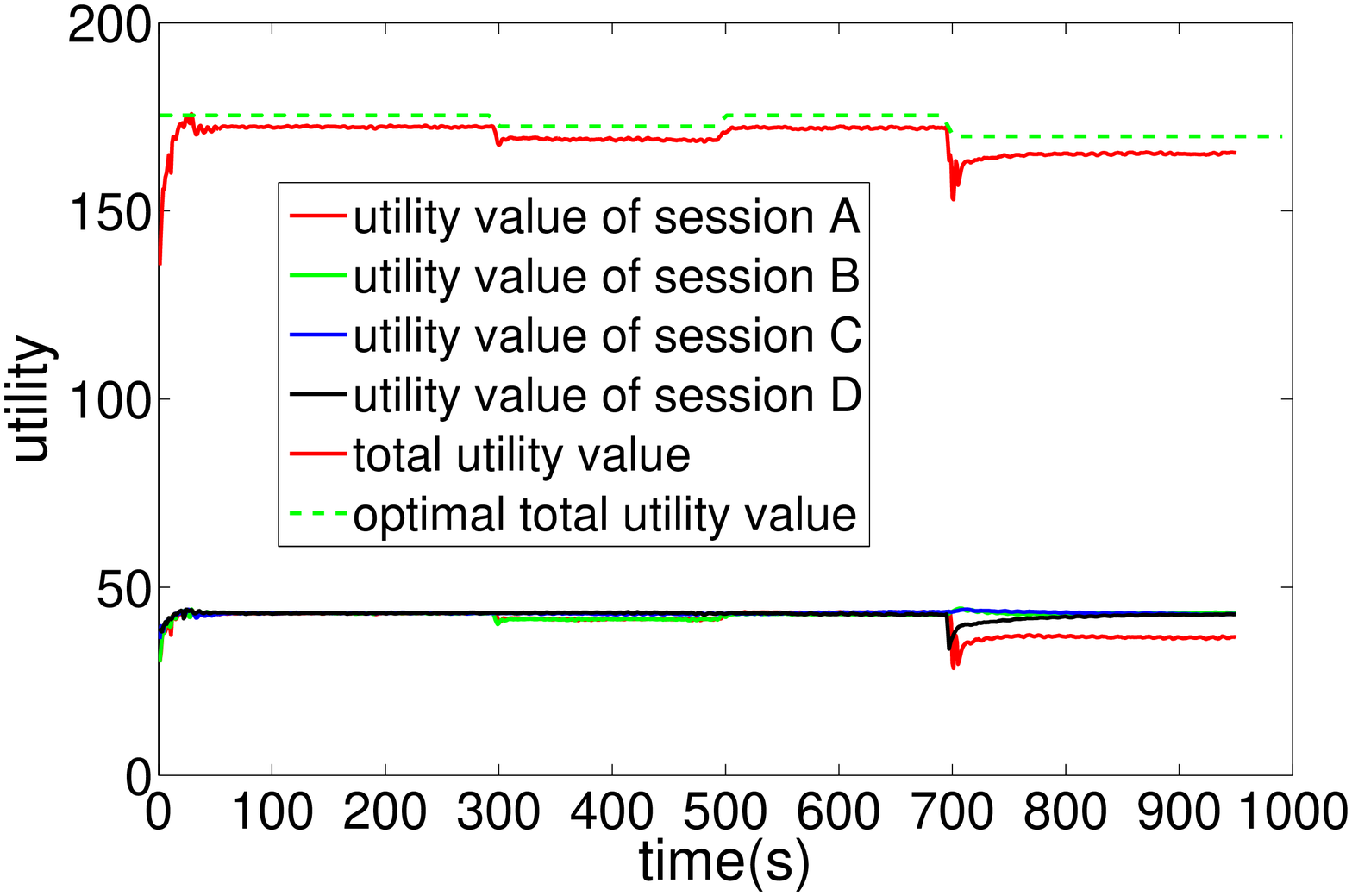}

\label{Fig:LAN_utility}}\subfloat[Average end-to-end delay and loss rate from node A to other nodes]{\includegraphics[width=6cm]{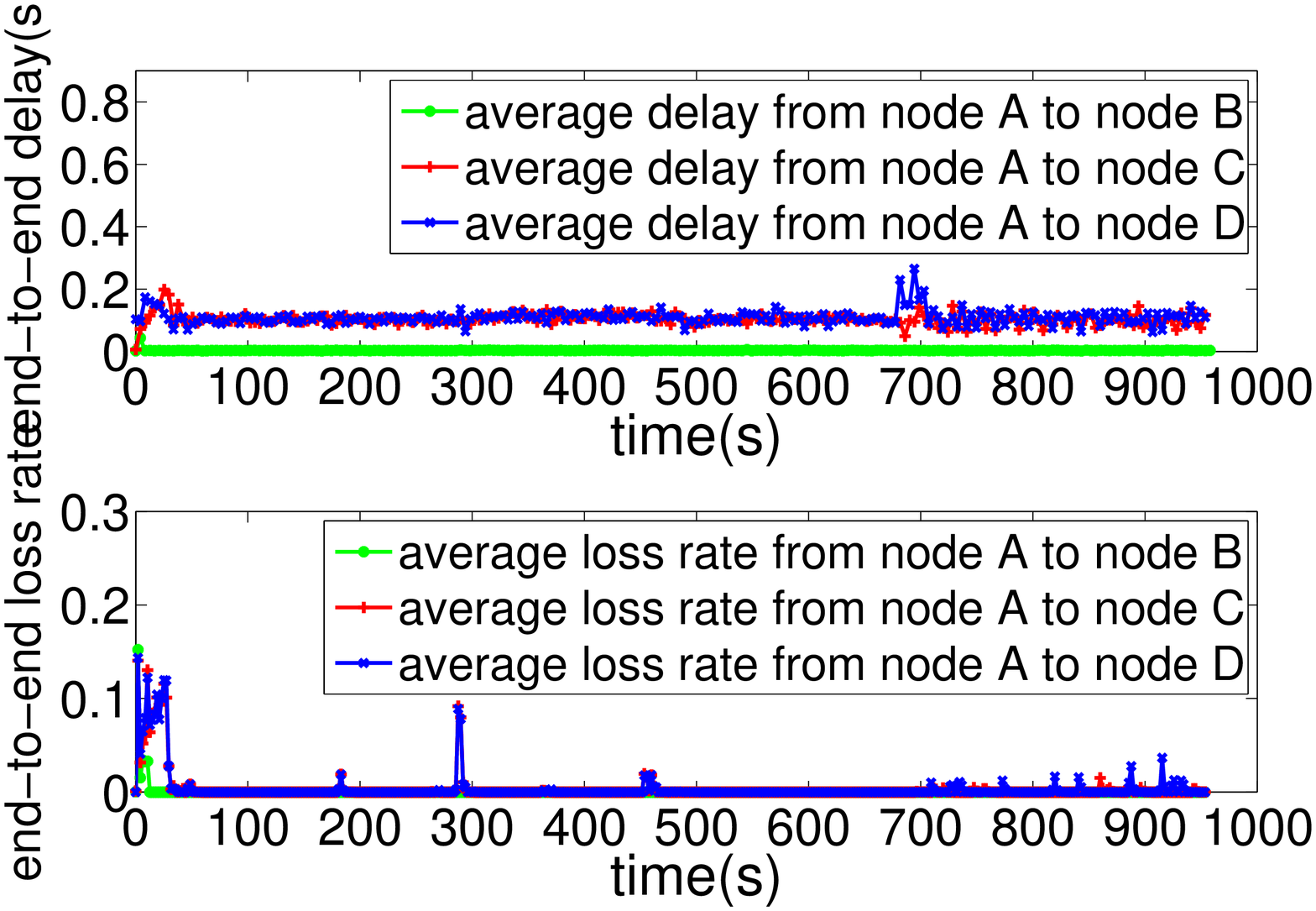}\label{Fig:LAN_loss_delay_A}}

\caption{Performance of \emph{Celerity} in the LAN Testbed Experiments. (a)-(d): Sending rates and receiving rates of individual sessions. (e): Utility value
achieved compared to the optimum. (f): End-to-end delay and loss rate of session $A$.\label{Fig:LAN_expr_results}}
\end{figure*}
\begin{figure*}[ht!]
\subfloat[Rate Performance of all Nodes]{\includegraphics[width=6cm]{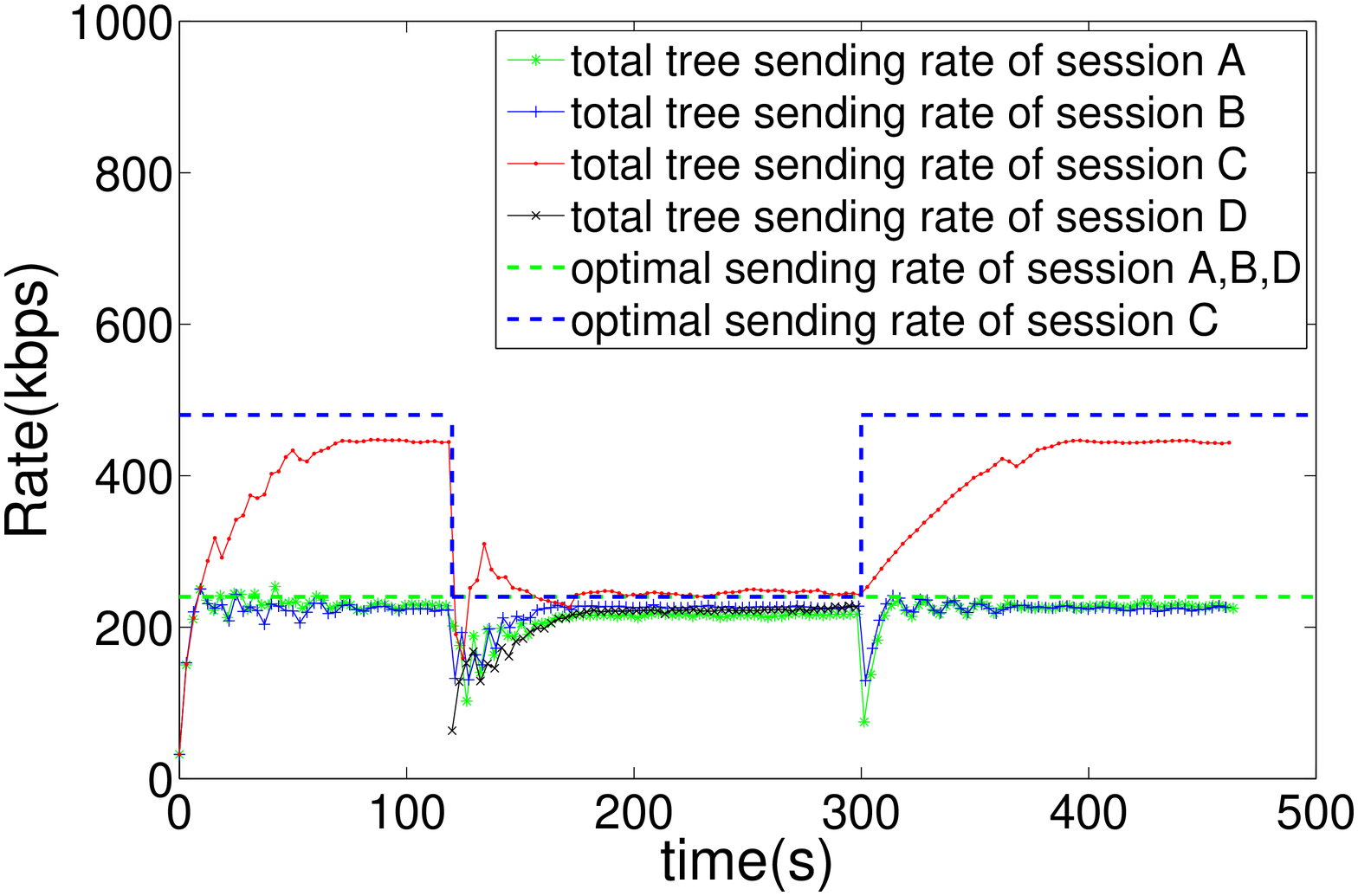}

\label{Fig:dyn_rates}}\subfloat[Average end-to-end delay and loss rate from node A to other nodes]{\includegraphics[width=6cm]{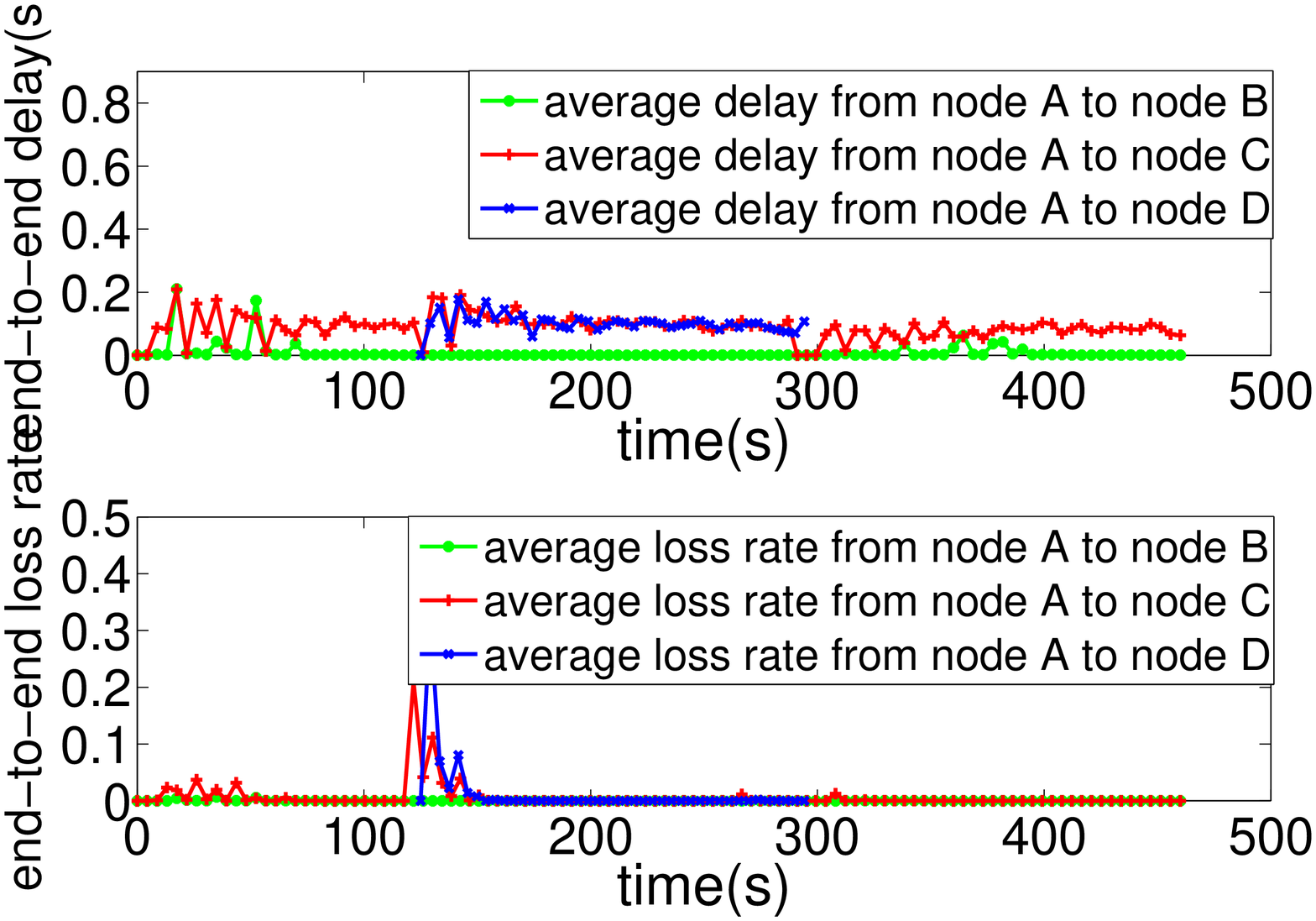}

\label{Fig:dyn loss delay A}}\subfloat[Average end-to-end delay and loss rate from node C to other nodes]{\includegraphics[width=6cm]{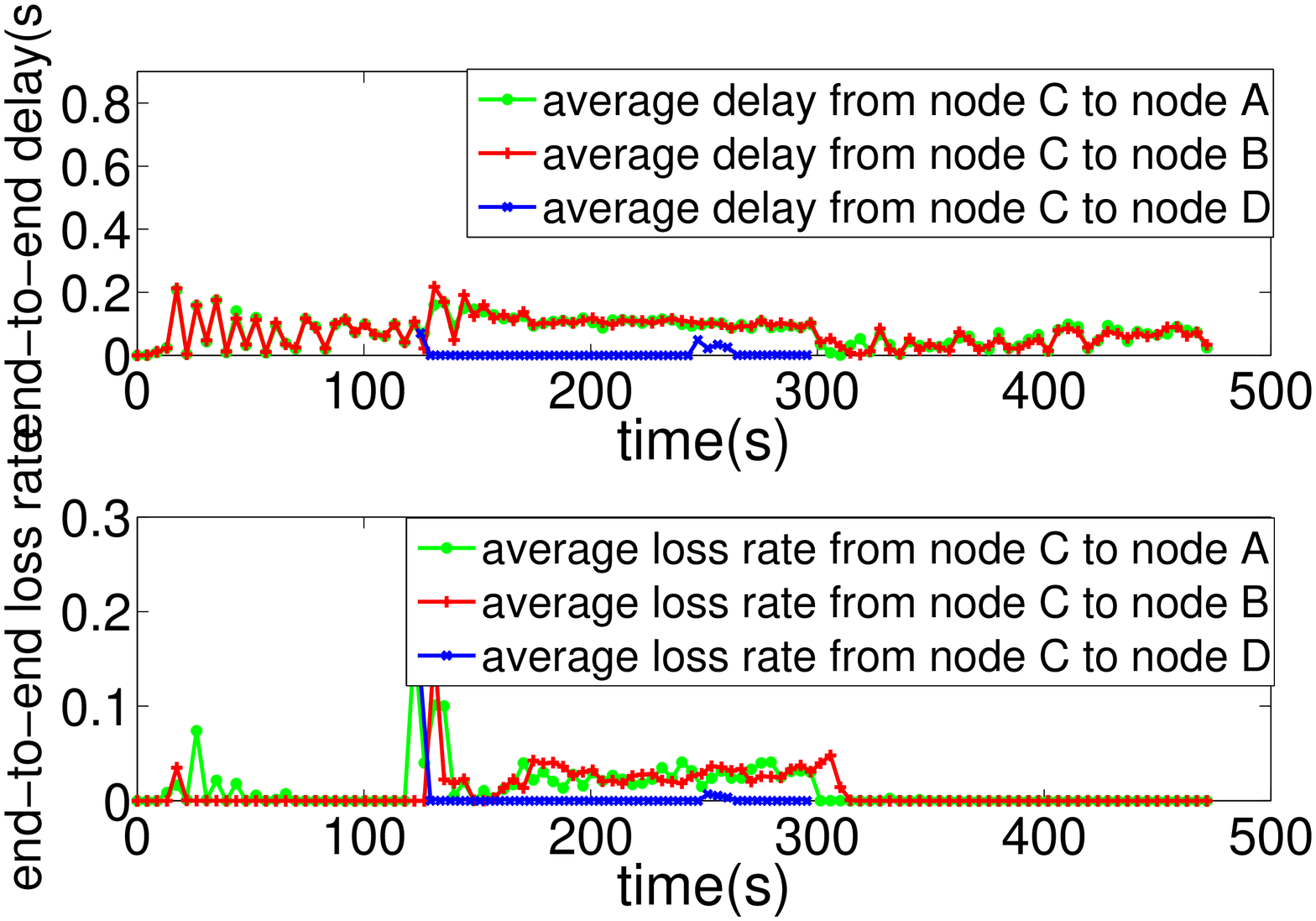}\label{Fig:dyn_loss_delay_C}}

\caption{Performance of \emph{Celerity} in the Peer Dynamics Experiments. (a)-(f): Sending rates of all sessions. (b)-(c): End-to-end delay and loss rate of session $A$ and $C$. \label{Fig:LAN_expr_results}}

\end{figure*}

\begin{figure*} [t]
\subfloat[Sending Rate of Node A (Hong Kong)]{\includegraphics[width=6cm]{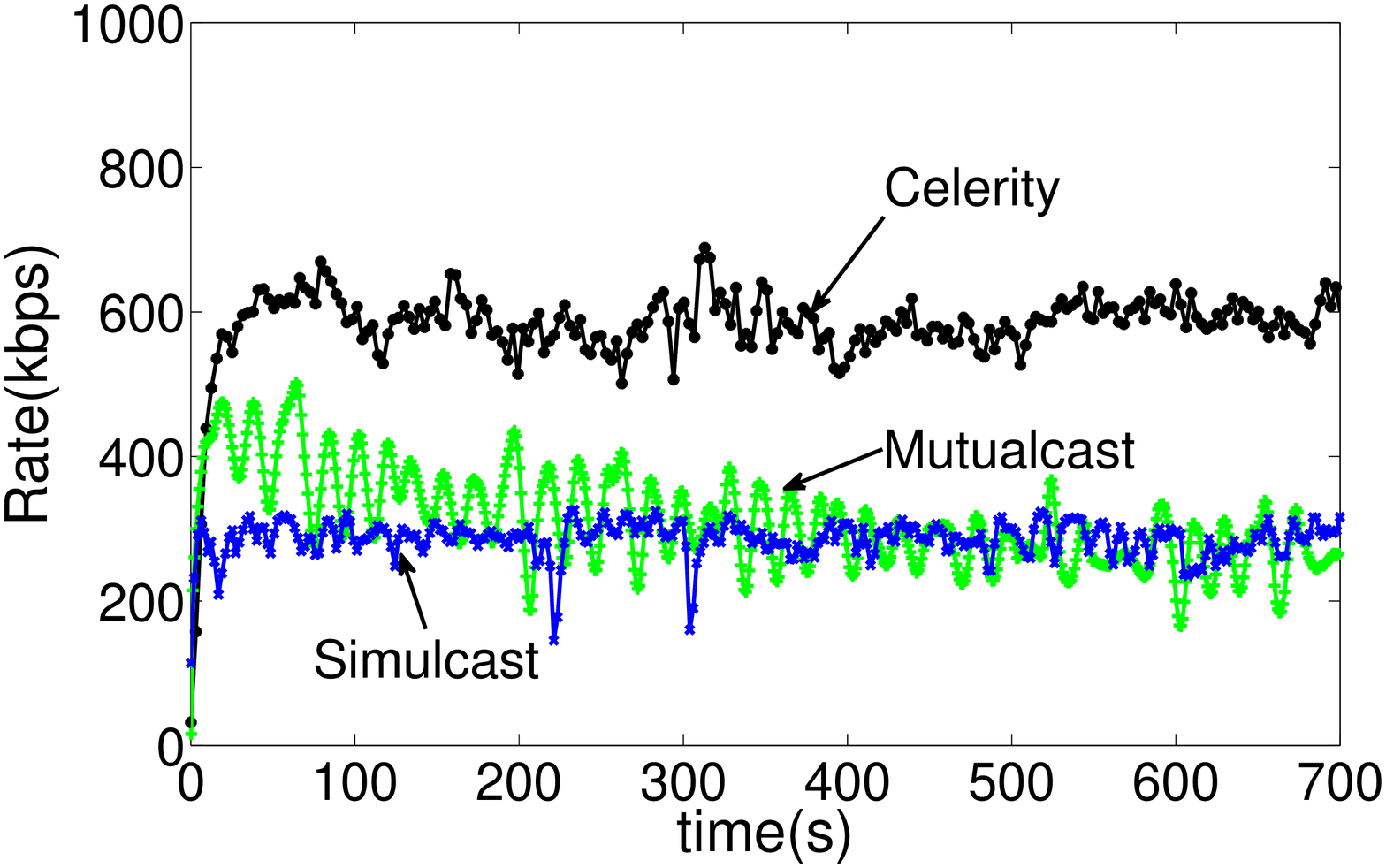}\label{Flo:fig.1}}\subfloat[Sending Rate of Node B (Hong Kong)]{\includegraphics[width=6cm]{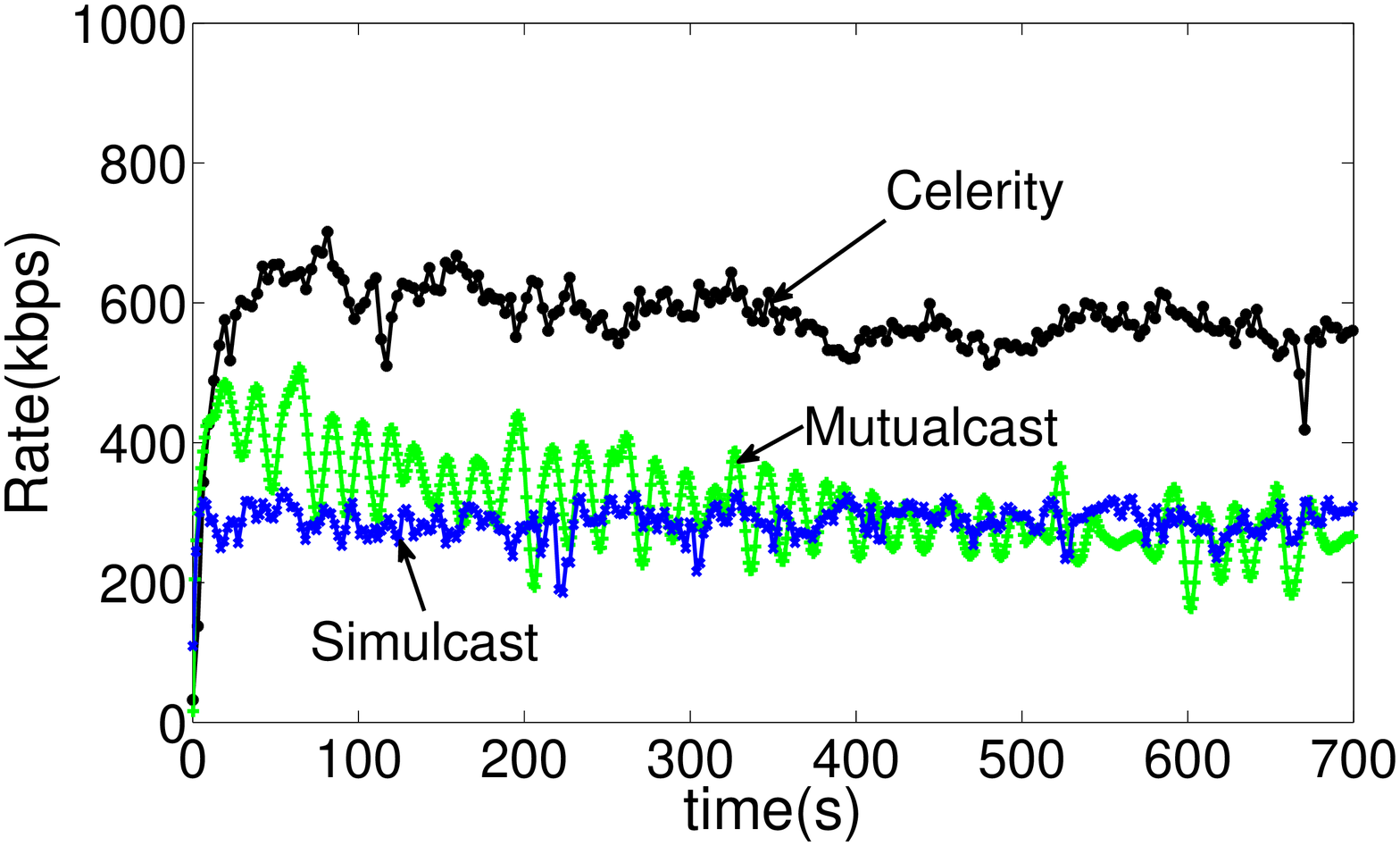}\label{Flo:fig.2}}\subfloat[Sending Rate of Node C (Redmond)]{\includegraphics[width=6cm]{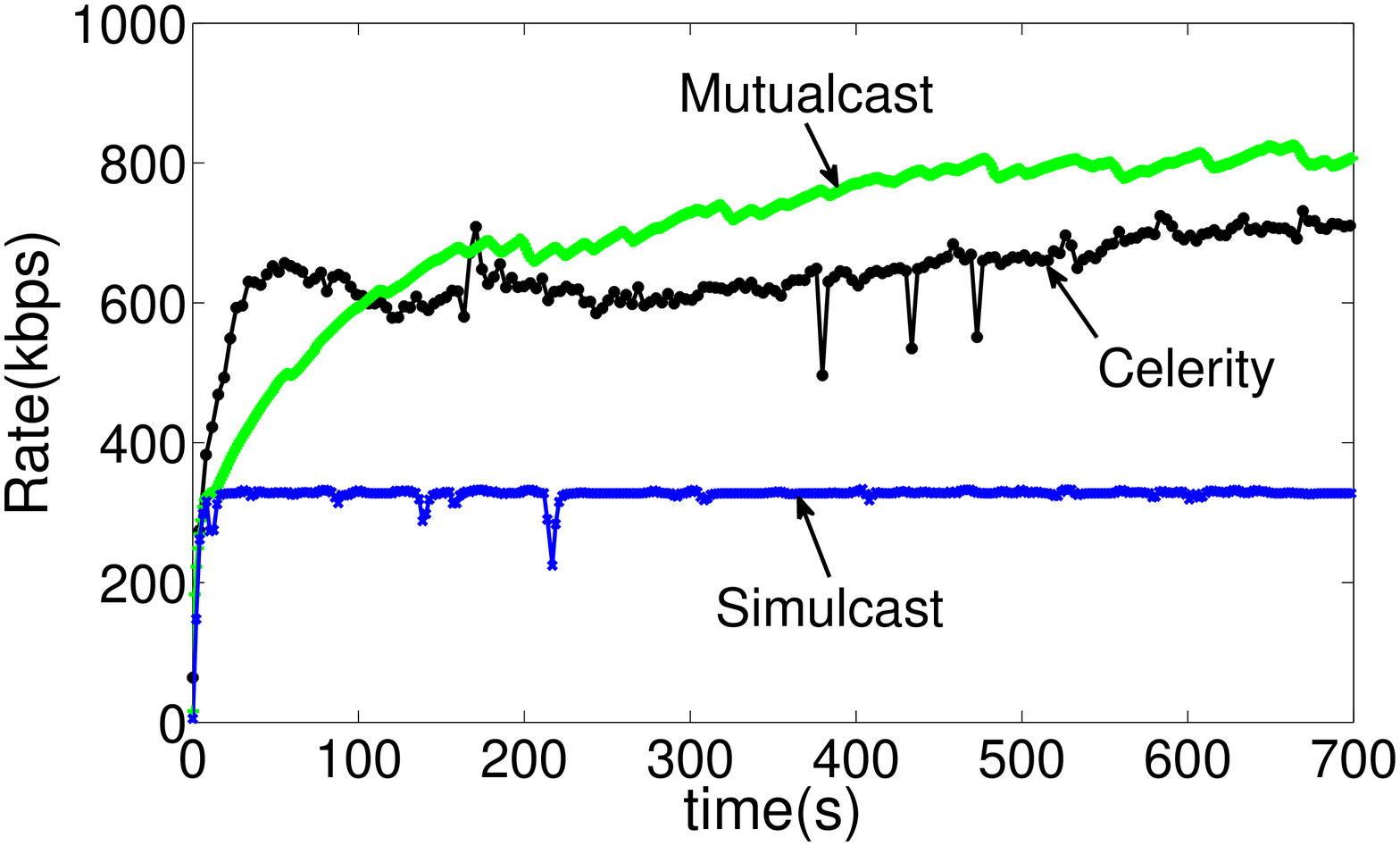}\label{Flo:fig.3}}

\subfloat[Sending Rate of Node D (Toronto)]{\includegraphics[width=6cm]{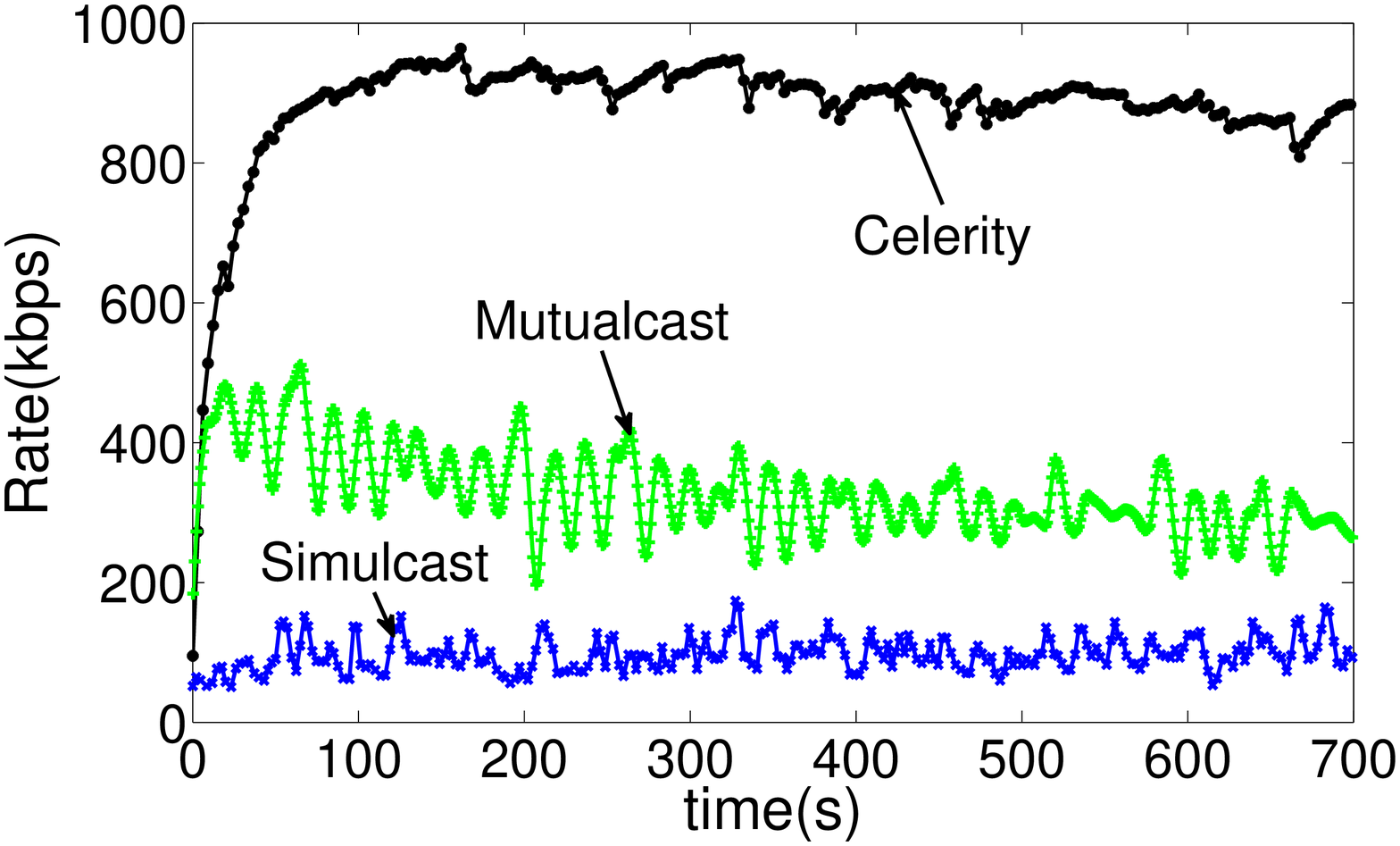}
\label{Flo:fig.4}}\subfloat[Total sending rate of all sessions]{\includegraphics[width=6cm]{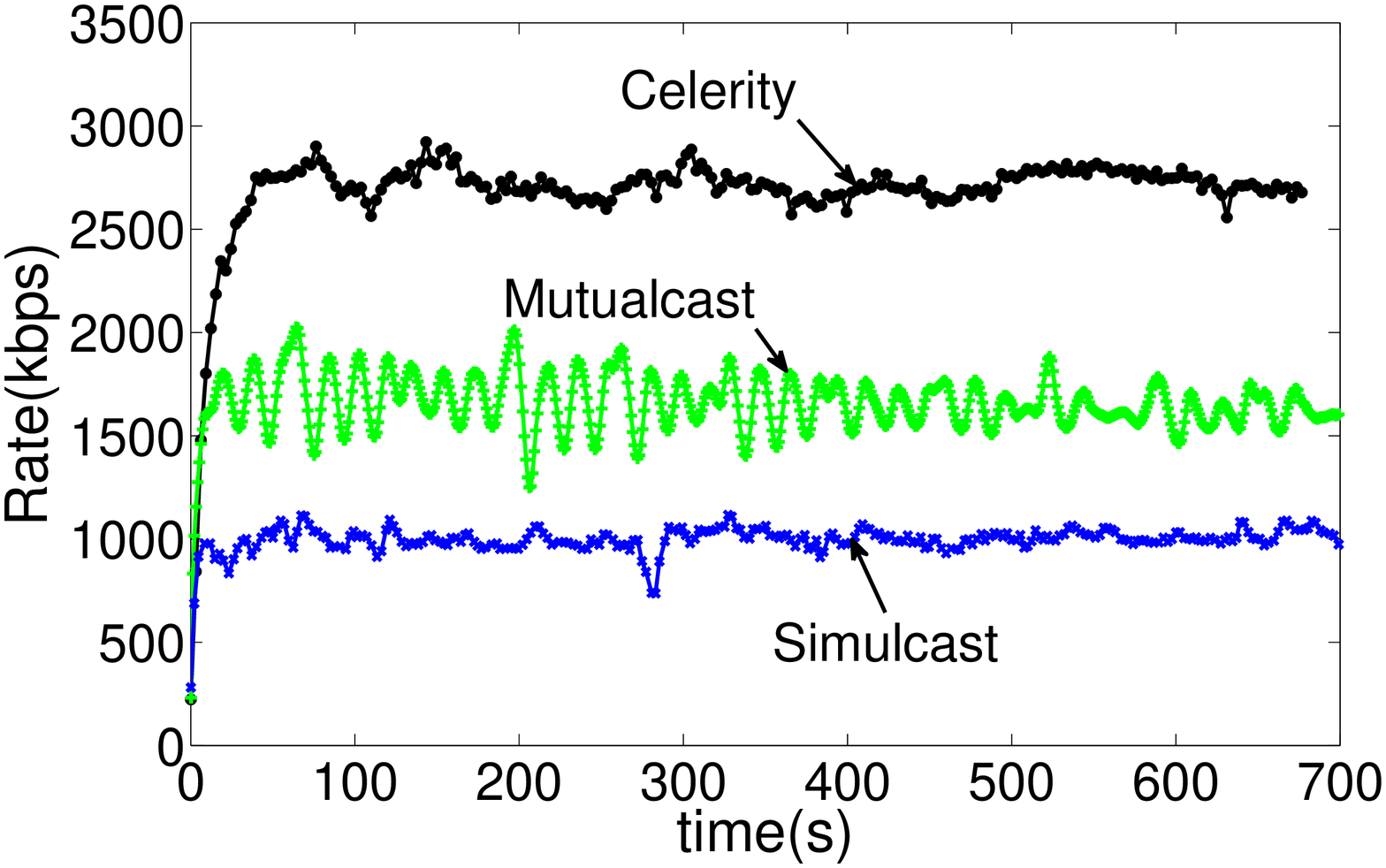}\label{Flo:fig.5}}\subfloat[Total utility of all sessions]{\includegraphics[width=6cm]{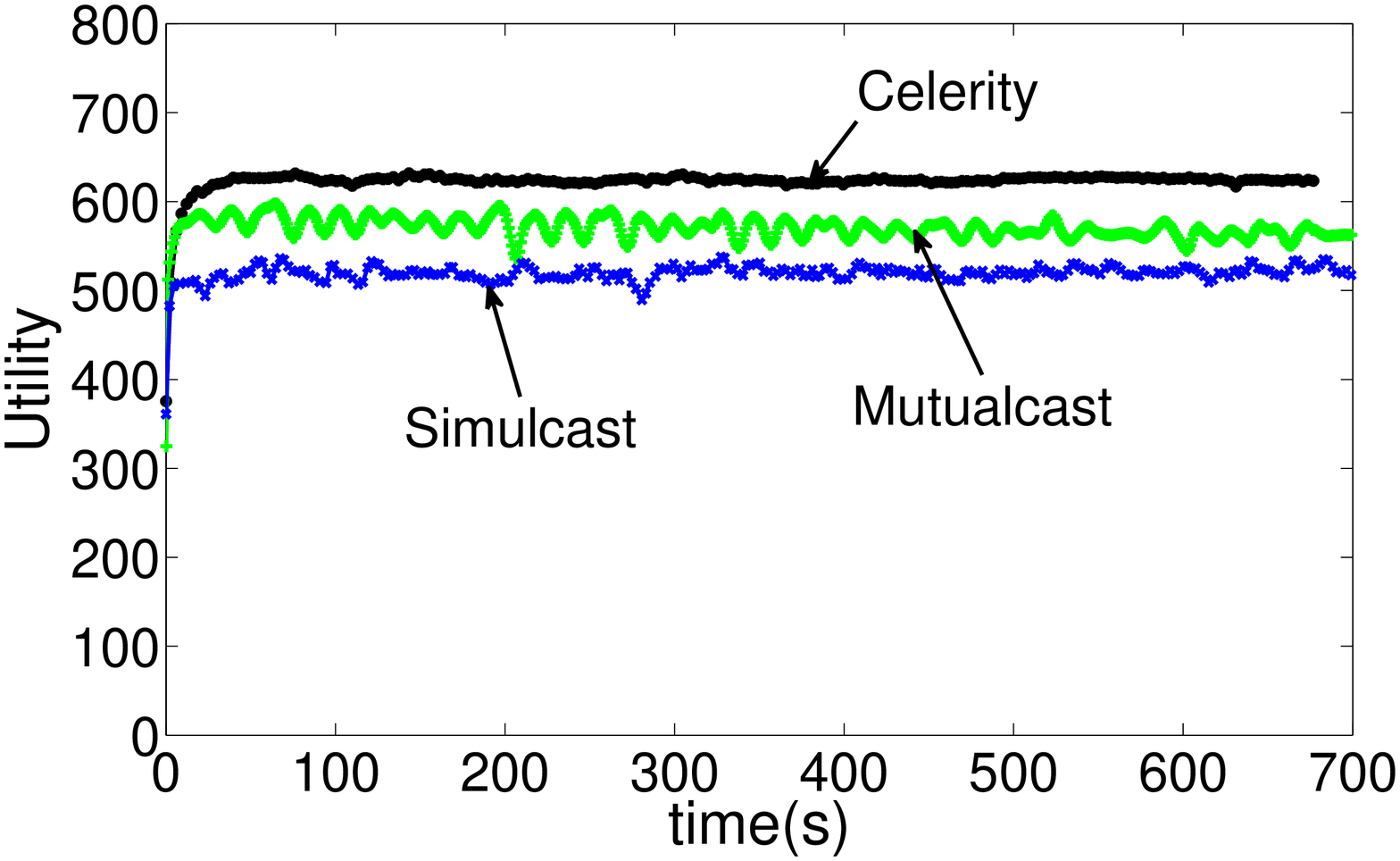}

\label{Flo:fig.6}}

\subfloat[Average end-to-end delay and loss rate from node A to other nodes]{\includegraphics[width=6cm]{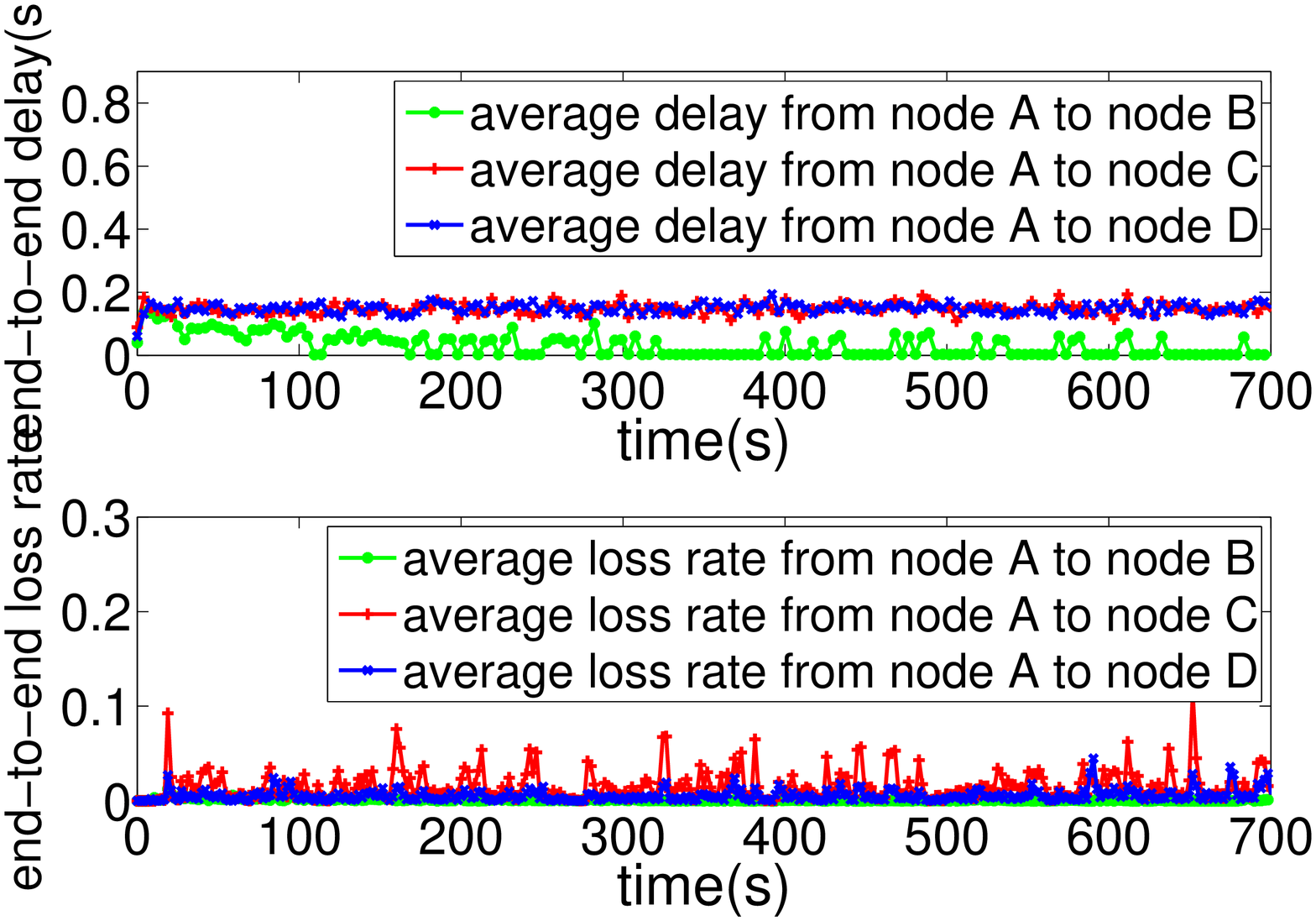}\label{Flo:fig.7}}\subfloat[Average end-to-end delay and loss rate from node C to other nodes]{\includegraphics[width=6cm]{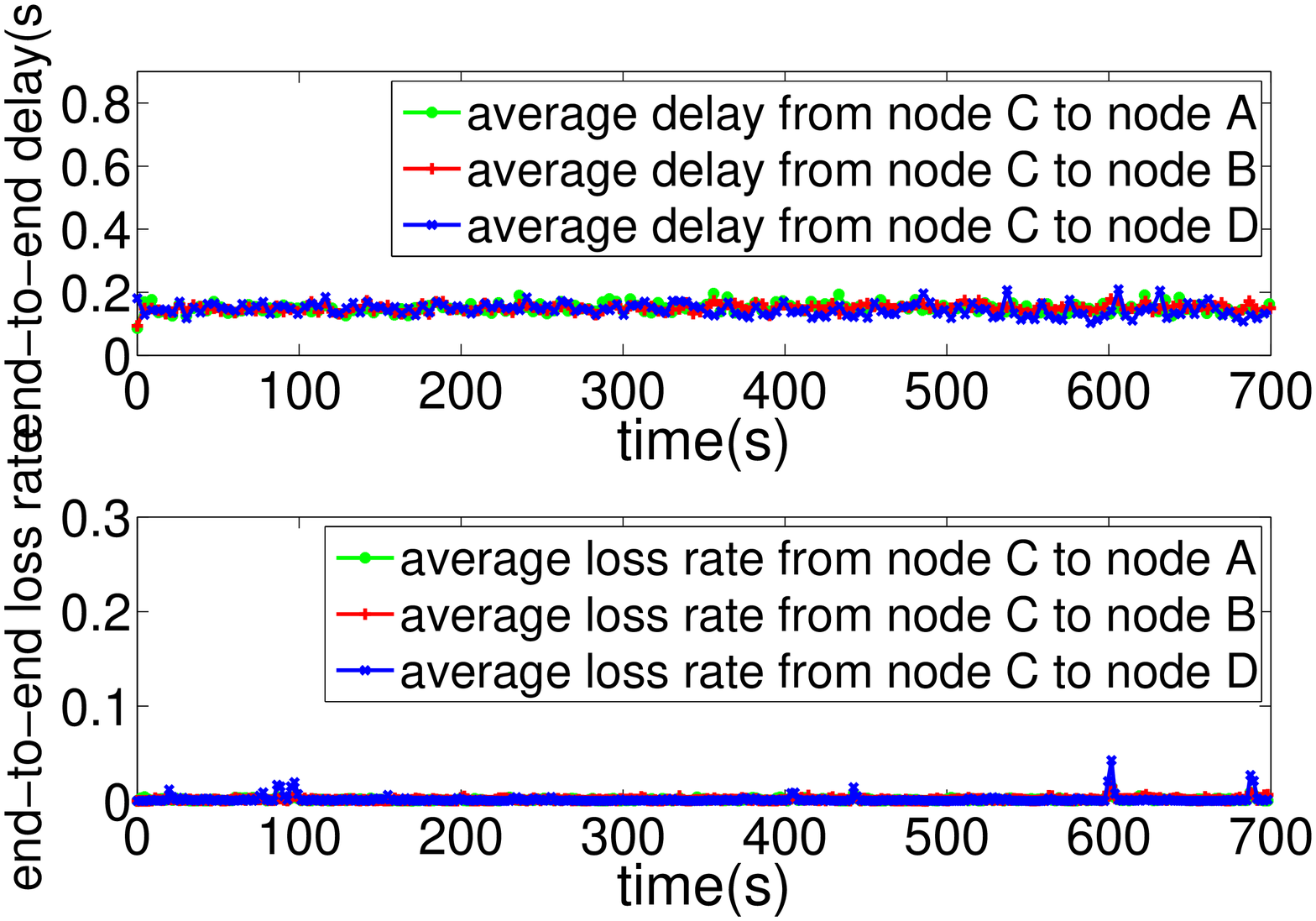}\label{Flo:fig.8}}\subfloat[Average end-to-end delay and loss rate from node D to other nodes]{\includegraphics[width=6cm]{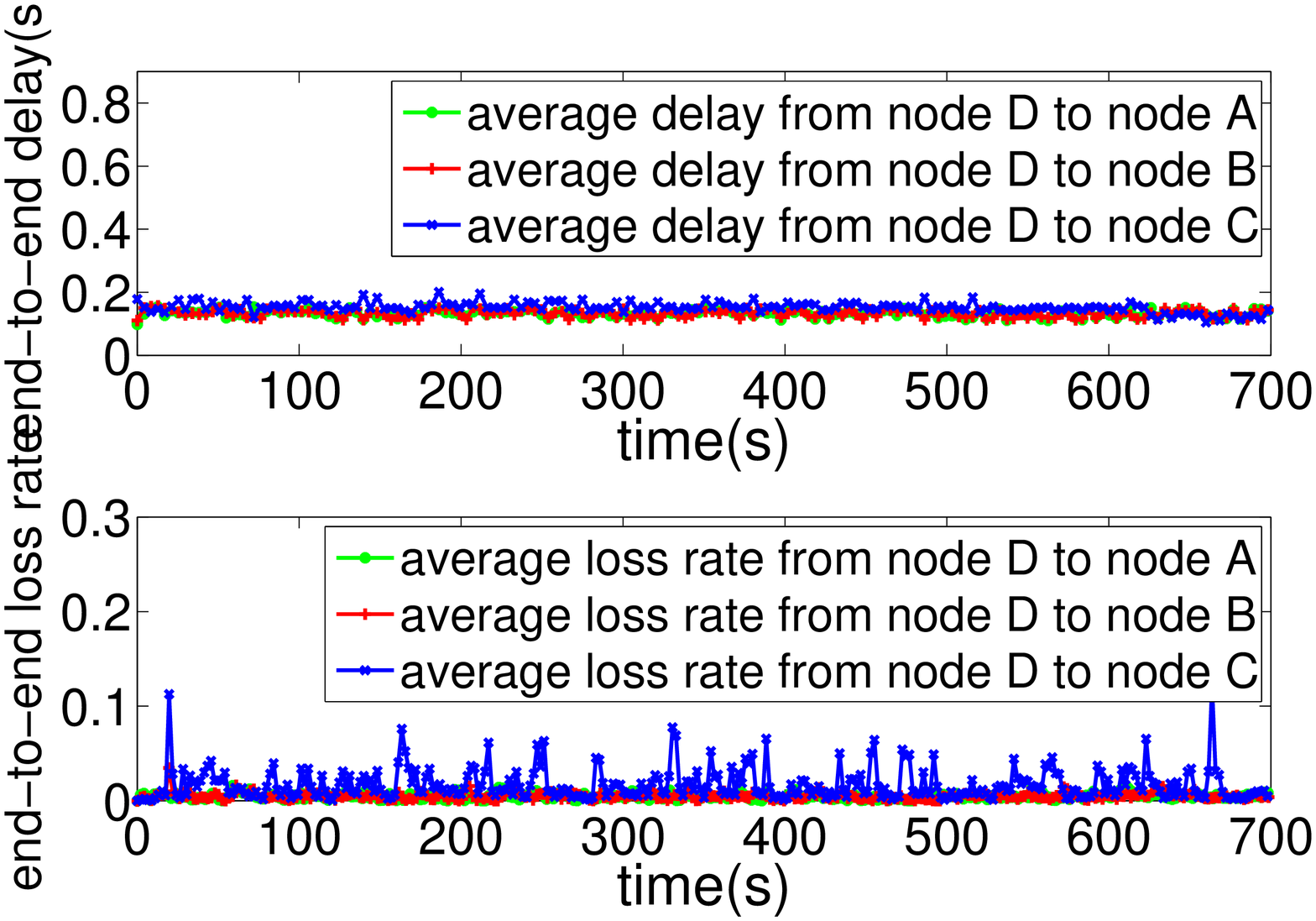}
\label{Flo:fig.9}}
\caption{Performance of four-party conferences over the Internet, running prototype
systems of \emph{Celerity}, Simulcast, and the scheme in \cite{chen2008ump}.
(a)-(d): Throughput of individual sessions. (e): Total throughput
of all sessions. (f): Utility achieved by different systems. (g)-(h):
End-to-end delay and loss rate of session $A$, $C$, and $D$ for the \emph{Celerity}
system.\label{fig:inet_results}}
\end{figure*}

Figs. \ref{Fig:LAN_rate_A}-\ref{Fig:LAN_rate_D} show the sending
rate of each session (one session originates from one node to all
other three nodes). For comparison, we also show the maximum achievable
rates by Simulcast and Mutualcast, as well as the optimal sending
rate of each session calculated by solving the problem in (\ref{eq:NUM1})-(\ref{eq:NUM2})
using a central solver. Fig. \ref{Fig:LAN_utility} shows the utility
obtained by \emph{Celerity} and its comparison to the optimal. Fig.
\ref{Fig:LAN_loss_delay_A} shows the average end-to-end delay and
packet loss rate of session $A$. Delay and loss performance of other
sessions are similar to those of session $A$.

In the following, we explain the results according to three different
experiment stages.

\begin{figure}[h!]
 \centering \includegraphics[width=7.4cm]{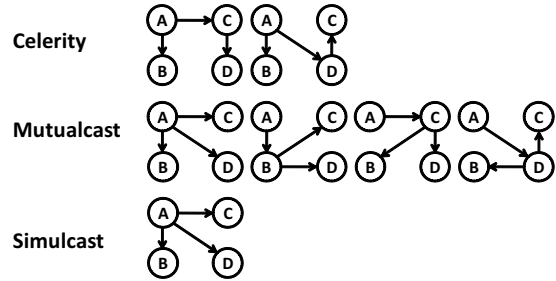} \caption{Session $A$'s trees used by \emph{Celerity} (upon convergence), Mutualcast
and Simulcast in the dumbbell topology, in the absence of network
dynamics.}

\label{Fig:trees}
\end{figure}

\subsubsection{Absence of Network Dynamics}

We first look at the first 300 seconds when there is no cross traffic
or link failure. In this time period, the experimental settings are
symmetric for all participating peers; thus the optimal sending rate
for each session is $240$ kbps.

As seen in Figs. \ref{Fig:LAN_rate_A}-\ref{Fig:LAN_rate_D}, \emph{Celerity}
demonstrates fast convergence: the sending rate of each session quickly
ramps up to $95$\% to the optimal within $50$ seconds. Fig. \ref{Fig:LAN_utility}
shows that \emph{Celerity} quickly achieves a close-to-optimal utility.
These observations indicate any other solution can at most outperform
\emph{Celerity} by a small margin.

As a comparison, we also plot the theoretical maximum rates achievable
by Simulcast and Mutualcast in Figs. \ref{Fig:LAN_rate_A}-\ref{Fig:LAN_rate_D}.
We observe that within $20$ seconds, our system already outperforms
the maximum rates of Simulcast and Mutualcast.

Upon convergence, \emph{Celerity} achieves sending rates that nearly
double the maximum rate achievable by Simulcast and Mutualcast. This
significant gain is due to that \emph{Celerity} can utilize the bottleneck
resource more efficiently, as explained below.

In Fig. \ref{Fig:trees}, we show the trees for session $A$ that
are used by \emph{Celerity}, Mutualcast and Simulcast in the dumbbell
topology. As seen, Simulcast and Mutualcast only explore 2-hop trees
satisfying certain structure, limiting their capability of utilizing
network capacity efficiently. In particular, their trees consumes
the bottleneck link resource twice, thus to deliver one-bit of information
it consumes two-bit of bottleneck link capacity. For instance, the
tree used by Simulcast has two branches $A\rightarrow C$ and $A\rightarrow D$
passing through the bottleneck links between $E$ and $F$, consuming
twice the critical resource. Consequently, the maximum achievable
rates of Simulcast and Mutualcast are all $120$ kbps. In contrast,
\emph{Celerity} explores all 2-hop delay-bounded trees, and upon convergence
utilizes the trees that only consume bottleneck link bandwidth once,
achieving rates that are close to the optimal of $240$ kbps.

Fig. \ref{Fig:LAN_loss_delay_A} shows the average end-to-end delay
and packet loss rate of session $A$. As seen, the packet loss rate
and delay are high initially, but decreases and stabilizes to small
values afterwards. The initial high loss rate is because \emph{Celerity}
increases the sending rates aggressively during the conference initialization
stage, in order to bootstrap the conference and explore network resource
limits. \emph{Celerity} quickly learns and adapts to the network topology,
ending up with using cost-effective trees to deliver data. After the
initialization stage, \emph{Celerity} adapts and converges gradually,
avoiding unnecessary performance fluctuation that deteriorates user
experience. By adapting to both delay and loss, we achieve low loss
rate upon convergence as compared to the case when only loss is taken
into account~\cite{Celerity2011nossdav}.

\subsubsection{Cross Traffic}

Between the 300th second and the 500th second, we introduce an 80kpbs
cross-traffic from node $E$ to node $F$. Consequently, the available
bottleneck bandwidth between $E$ and $F$ decreases from 480 kbps
to 400 kbps. We calculate the optimal sending rates during this time
period to be $200$ kbps for sessions $A$ and $B$, and remain $240$
kbps for sessions $C$ and $D$.

As seen in Figs. \ref{Fig:LAN_rate_A}-\ref{Fig:LAN_rate_D}, \emph{Celerity}
quickly adapts to the bottleneck bandwidth reduction. \emph{Celerity}'s
adaptation is expected from its design, which infers from loss and
delay the available resource and adapt accordingly. From Fig. \ref{Fig:LAN_loss_delay_A},
we can see a spike in session $A$'s packet loss rate around 300th
second, at which time the available bottleneck bandwidth reduces.
The link rate control modules in \emph{Celerity} senses this increased
loss rate, adjusts, and reports the reduced (overlay) link rates to
node $A$. Upon receiving the reports, the tree-packing module in
\emph{Celerity} adjusts the source sending rate accordingly, adapting
the system to a new close-to-optimal operating point. At 500th second,
the cross traffic is removed and the available bottleneck bandwidth
between $E$ and $F$ restores to 480kbps. \emph{Celerity} also quickly
learns this change and adapts to operate at the original point, evident
in Figs. \ref{Fig:LAN_rate_A}-\ref{Fig:LAN_rate_B}.

\subsubsection{Link Failure}

Between the 700th second and the 1000th second, we disconnect the
physical link between $A$ and $E$. Consequently, node $A$ cannot
use the $2$-hop threes with node $C$ ($D$) being intermediate nodes;
similarly node $C$ ($D$) cannot use the $2$-hop threes with node
$A$ being intermediate nodes. They can, however, still use the trees
with node $B$ as intermediate nodes. We compute the theoretical optimal
sending rates during this time period to be $240$ kbps for all sessions.

We observe from Fig. \ref{Fig:LAN_rate_A} that node $A$'s sending
rate first drops immediately upon link failure, then quickly adapts
to the new operating point of around $120$ kbps, only half of the
theoretical optimal. This is because \emph{Celerity} only explores
$2$-hop trees for content delivery while in this case $3$-hop trees
(e.g., $A\rightarrow B\rightarrow C\rightarrow D$) are needed to
achieve the optimal. It is of great interest to explore source rate
control mechanisms beyond this $2$-hop tree-packing limitation to
further improve the performance without incurring excessive overhead.

In Figs.~\ref{Fig:LAN_rate_D}, we observe the sending rate of session
$D$ first drops and then climbs back. This is because session $D$
\emph{happens} to use the trees with node $A$ being intermediate
nodes right before the link failure. The link failure breaks session
$D$'s trees, thus session $D$'s rate drops dramatically. \emph{Celerity}
detects the significant change and adapts to use the trees with $B$
as intermediate nodes for session $D$. Session $D$'s rate thus gradually
restore to around the optimal. These observations show the excellent
adaptability of \emph{Celerity} to abrupt network condition changes.

As a comparison, we observe that Simulcast's maximum achievable rates
of session $A$, $C$, and $D$ all drop to zero upon the link failure.
This is because there is no direct overlay link between $A$ and $C$
($D$) after the link failure. Consequently, Simulcast is not able
to broadcast the source's content to all the receivers in these sessions,
resulting in zero session rates.
\subsection{Peer Dynamics Experiments\label{ssec:expr:peer dynamic}}

In order to evaluate the \emph{Celerity} performance in peer dynamics scenario, we conduct another experiment over the same LAN testbed in Fig. \ref{Fig:topology}. We first run a three-party conference among node $A$, $B$, and $C$, at the 120th second, a node $D$ joins the conferencing session and leaves at the 300th second, the entire conferencing session lasts for 480 seconds.

Fig. \ref{Fig:dyn_rates} shows the sending rate of each session as well as the optimal sending rate of each session, Fig. \ref{Fig:dyn loss delay A}-\ref{Fig:dyn_loss_delay_C} show the average end-to-end delay and packet loss rate of session $A$ and $C$. Delay and loss performance of session $B$ are similar to those of session $A$.

As seen in Fig. \ref{Fig:dyn_rates}, when node $D$ joins the conferencing session at the 120th second, the sending rates of session $A$, $B$ and $C$ first drop immediately, then quickly adapt to close to the optimal value again. This is because when node $D$ joins, the initial allocated rates for each session in the overlay links from other nodes to node $D$ are very low, when node $A$, $B$ and $C$ pack trees respectively according to the allocated rates to deliver their data to the receivers including node $D$, the achieved sending rates are low. Then, \emph{Celerity} detects the change of underlay topology, updates the allocated rates and quickly converges to the new close to optimal operating point. When node $D$ leaves, we also observe that \emph{Celerity} quickly adapts to the peer dynamic.

\emph{Celerity's} excellent performance adapting to peer dynamics is expected from its design. We involve both loss and queuing delay in our design, when peers join and leave, loss and queuing delay reflect such events well, thus allowing \emph{Celerity} to adapt rapidly to the peer dynamics. For instance, in this experiment when node $D$ joins the conferencing session, we observe a spike in session $A$'s end-to-end delay and packet loss rate in Fig. \ref{Fig:dyn loss delay A}.

In Fig. \ref{Fig:dyn_rates} another important observation is that as compared to the conference initialization stage, the convergence speed of node $C$ after node $D$ leaves the conferencing session is slow. This is because during the conference initialization stage, \emph{Celerity} uses a method called "quick start" described in Section \ref{ssec:fastBootstrapping} to quickly ramp up the rates of all sessions, while after the initialization stage, such method is not used in order to avoid unnecessary performance fluctuation. It is of great interest to design source rate control mechanisms to achieve quick convergence in peer dynamics scenario without incurring system fluctuation.

\subsection{Internet Experiments\label{ssec:expr:internet}}

Beside the prototype \emph{Celerity} system, we also implement two
prototype systems of Simulcast and Mutualcast, respectively. Both
\emph{Celerity} and Mutualcast use the same log utility functions
in their rate control modules. We evaluate the performance of these
systems in a four-party conferencing scenario over the Internet.

We use four PC nodes that spread two continents and tree countries
to form the conferencing scenario. Two of the PC nodes are in Hong
Kong, one is in Redmond, Washington, US, and the last one is in Toronto,
Canada. This setting represents a common global multi-party conferencing
scenario.

We run multiple 15-minute four-party conferences using the prototype
systems, in a one-by-one and interleaving manner. We select one representative
run for each system, and summarize their performance in Fig. \ref{fig:inet_results}.

Figs. \ref{Flo:fig.1}-\ref{Flo:fig.4} show the rate performance
of each session. (Recall that a session originates from one node to
all other three nodes.) As seen, all the session rates in \emph{Celerity}
quickly ramp up to near-stable values within 50 seconds, and outperforms
Simulcast within 10 seconds. Upon stabilization, \emph{Celerity} achieves
the best throughput performance among the three systems and Simulcast
is the worst. For instance, all the session rates in \emph{Celerity}
is 2x of those in Simulcast and Mutualcast, except in session C where
Mutualcast is able to achieve a higher rate than \emph{Celerity}.

We further observe \emph{Celerity}'s superior performance in Fig.
\ref{Flo:fig.5}, which shows the aggregate session rates, and in
Fig. \ref{Flo:fig.6}, which shows the total achieved utilities. In
both statistics, \emph{Celerity} outperforms the other two systems
by a significant margin. Specifically, the aggregate session rate
achieved by \emph{Celerity} is 2.5x of that achieved by Simulcast,
and is 1.8x of that achieved by Mutualcast.

These results show that our theory-inspired \emph{Celerity} solution
can allocate the available network resource to best optimize the system
performance. Mutualcast aims at similar objective but only works the
best in scenarios where bandwidth bottlenecks reside only at the edge
of the network \cite{chen2008ump}.

Figs. \ref{Flo:fig.7}-\ref{Flo:fig.9} show the average end-to-end
loss rate and delay from source to receivers for session $A$, session
$C$ and session $D$. The results for session $B$ is very similar
to session $A$ and is not included here. As seen, the average end-to-end
delays of all sessions are within 200 ms, which is our preset delay
bound for effective interactive conferencing experience. The average
end-to-end loss rate for all sessions are at most 1\%-2\% upon system
stabilization.

The overall operation overhead of \textit{Celerity}
in the 4-party Internet experiment is around 3.9\%. In particular,
the packet overhead accounts for 3.4\%, and the link-rate control and
link-state report overhead is around 0.5\%.

\section{Concluding Remarks}
\label{sec:conclusion}

With the proliferation of front-facing cameras on mobile devices, multi-party video conferencing will soon become an utility that both businesses and consumers would find useful.  With {\em Celerity,} we attempt to bridge the long-standing gap between the bit rate of a video source and the highest possible delay-bounded broadcasting rate that can be accommodated by the Internet where \emph{the bandwidth bottlenecks can be anywhere in the network}.  This paper reports {\em Celerity} solution as a first step in making this vision a reality: by combining a polynomial-time tree packing algorithm on the source and an adaptive rate control along each overlay link, we are able to maximize the source rates without any {\em a priori} knowledge of the underlying physical topology in the Internet.  {\em Celerity} has been implemented in a prototype system, and extensive experimental results in a ``tough'' dumbbell LAN testbed and on the Internet demonstrate {\em Celerity}'s superior performance over the state-of-the-art solution Simulcast and Mutualcast.

As future work, we plan to explore source rate control mechanisms beyond the $2$-hop tree-packing limitation in {\em Celerity} to further improve its performance without incurring excessive overhead.

\section*{APPENDIX}
\begin{description}
\item [{{{{{{{{{{{{A.}}}}}}}}}}}}] Proof of Theorem
$2$
\end{description}
\textit{Proof:} Firstly, we prove the minimum of the min-cuts separating
the source and receivers in $\mathcal{D}_{m}$ can be expressed as

\[
R_{m}(\boldsymbol{c}_{m},D)=\underset{j}{\min}\underset{v\in\{r_{i}\}\cup\{h_{k}\}}{\sum}\min\left\{ c_{m,s\rightarrow v},c_{m,v\rightarrow t_{j}}\right\}
\]
 .

In the overlay graph $\mathcal{D}_{m}$, the minimum of the min-cuts
is $\min_{t_{j}\epsilon T}$ $MinCut\left(s,\, t_{j}\right)$. where
$T$ is the set of receivers, and $MinCut\left(s,\, t_{j}\right)$
is the min-cut separating the source $s$ and receiver $t_{j}$. The
min-cut separating the source and a receiver can be achieved by finding
the maximum unit-capacity disjoint paths from the source to the receiver.
The structure of the graph $\mathcal{D}_{m}$ is so special that for
each receiver $t_{j}$ we can compute the maximum number of edge-disjoint
paths from $s$ to $t_{j}$ easily.

In the graph $\mathcal{D}_{m}$ we represent each edge with capacity
$m$ by $m$ parallel edges, each with unit capacity. For each receiver
node, say $t_{j}$, due to the special structure of the graph, we
can find these edge-disjoint paths in a very simple way. Since there
are only 2-hop paths in the graph $\mathcal{D}_{m}$, so a path from
$s$ to $t_{j}$ must go through one of the intermediate nodes. Thus
for each intermediate node, say $e$ , we can find $\min\left\{ c_{m,s\rightarrow e},\, c_{m,e\rightarrow t_{j}}\right\} $
edge-disjoint paths from $s$ to $e$ and then to $t_{j}$. Therefore,
we can have

\[
MinCut\left(s,\, t_{j}\right)=\underset{v\in\{r_{i}\}\cup\{h_{k}\}}{\sum}\min\left\{ c_{m,s\rightarrow v},c_{m,v\rightarrow t_{j}}\right\}
\]

Consequently, the minimum of the min-cuts separating the source and
receivers can be expressed as

\[
R_{m}(\boldsymbol{c}_{m},D)=\underset{j}{\min}\underset{v\in\{r_{i}\}\cup\{h_{k}\}}{\sum}\min\left\{ c_{m,s\rightarrow v},c_{m,v\rightarrow t_{j}}\right\}
\]
 .

Next, we prove the tree packing algorithm can achieve the minimum
of the min-cuts separating the source and receivers in the two layer
graph $\mathcal{D}_{m}$. This tree packing algorithm is developed
based on the Lovasz's constructive proof \cite{lovasz1976two} to
Edmonds' Theorem\cite{Edmonds1973}. To proceed, we firstly apply
the Lovasz's constructive proof to our two layer graph $\mathcal{D}_{m}$
and based on the proof, we can directly have the tree packing algorithm.

\textbf{Notations}: Let $G$ be a digraph with a source $a$. We assume
all edges have unit-capacity and allowing multiple edges for each
ordered node pair. $V(G)$ and $E(G)$ denote its vertex set and edge
set. A branching (rooted at $a$) is a tree which is directed in such
a way that each receiver $t_{i}$ has one edge coming in. A $cut$
of $G$ determined by a set $S\subset V(G)$ is the set of edges going
from $S$ to $V(G)-S$ and will be denoted by $\vartriangle_{G}(S)$,
we also set $\delta_{G}(S)=|\vartriangle_{G}(S)|$.

\textbf{Theorem:} In the two layer graph\textbf{ }$\mathcal{D}_{m}$,
if $\delta_{G}(S)\geq k$ for every $S\subset V(G),\, a\in S,\,\exists t_{i}\in V(G)-S$
then there are $k$ edge-disjoint branchings rooted at $a$.

\textit{Lovasz's constructive proof:} We use induction on $k$. It
is obvious that the theorem holds when $k=0$.

Let $F$ be a set of edges satisfying the following coditions

(i) $F$ is an arborescence rooted at $a$.

(\textbf{Definition}: In graph theory, an arborescence is a directed
graph in which, for a vertex $u$ called the root and any other vertex
$v$, there is exactly one directed path from $u$ to $v$. Equivalently,
an arborescence is a directed, rooted tree in which all edges point
away from the root. Every arborescence is a directed acyclic graph
(DAG), but not every DAG is an arborescence.)

(ii) $\delta_{G-F}(S)\geq k-1$ for every $S\subset V(G),\, a\in S,\,\exists t_{i}\in V(G)-S$.

If $F$ cover all receivers $t_{i}$, i.e., it is a branching then
we are finished: $G-F$ contains $k-1$ edge-disjoint branchings and
$F$ is in the $k$th one.

If $F$ only covers a set $T\subset V(G),$ which do not cover all
receivers, i.e., there exist some receivers $t_{i}\notin T$. We show
we can add an edge $e\in\vartriangle_{G}(T)$ to $F$ so that the
arising arborescence $F+e$ still satisfies (i) and (ii). Noting that
if $r_{i}\in T,$ then $t_{i}\in T,$ because there are infinite unit-capacity
edges from $r_{i}$ to $t_{i}$, adding a edge from $r_{i}$ to $t_{i}$
to $F$ can still satisfies (i) and (ii).

Consider a maximal set $A\subset V(G)$ such that

(a) $a\in A$;

(b) There is at least one receiver $t_{i}\notin A\cup T$;

(c) $\delta_{G-F}(A)=k-1$.

If no such $A$ exists any edge
\begin{eqnarray*}
e & \in & \{(r_{i},\, t_{j})|r_{i}\in T,\, t_{j}\in V(G)-T\}\\
 &  & \cup\{(h_{i},\, t_{j})|h_{i}\in T,\, t_{j}\in V(G)-T\}\\
 &  & \cup\{(a,\, r_{j})\}|t_{j}\notin T\}\cup\{(a,\, h_{i})|h_{i}\notin T\}
\end{eqnarray*}
 can be added to $F$.

Otherwise,

Since

\[
\delta_{G-F}(A\cup T)=\delta_{G}(A\cup T)\geq k,
\]

we have $A\cup T\neq A,$ $T\nsubseteq A$. Also,

\[
\delta_{G-F}(A\cup T)>\delta_{G-F}(A)
\]

and so, there must be an edge $e=(x,\, y)$ which belongs to $\vartriangle_{G-F}(A\cup T)-\vartriangle_{G-F}(A).$
Hence $x\in T-A$ and $y\in V(G)-T-A$. We claim $e$ can be added
to $F$, i.e., $F+e$ satisfies (i) and (ii).

Noting that due to the special structure of $\mathcal{D}_{m}$,

\begin{eqnarray*}
e & = & (x,\, y)\in\{(r_{i},\, t_{j})|r_{i}\in T-A,\, t_{j}\in V(G)-T-A\}\\
 &  & \cup\{(h_{i},\, t_{j})|h_{i}\in T-A,\, t_{j}\in V(G)-T-A\}
\end{eqnarray*}

So $y$ must be a receiver.

It is obvious that $F+e$ still satisfies (i) .

Let $S\subset V(G),\, a\in S,\,\exists t_{i}\in V(G)-S$. If $e\notin\vartriangle_{G-F}(S)$
then

\[
\delta_{G-F-e}(S)=\delta_{G-F}(S)\geq k-1.
\]

If $e\in\vartriangle_{G-F}(S)$ then $x\in S,\, y\in V(G)-S$. We
use the inequality

\begin{equation}
\delta_{G-F}(S\cup A)+\delta_{G-F}(S\cap A)\leq\delta_{G-F}(S)+\delta_{G-F}(A)\label{eq:inequ_1-1-1}
\end{equation}

which follows by an easy counting.

Since $a\in S\cap A,$ and there exist a receiver $y\in V(G)-S\cap A$,
we have

\[
\delta_{G-F}(A)=k-1,\qquad\delta_{G-F}(S\cap A)\geq k-1,
\]

and by the maximality of $A$,

\[
\delta_{G-F}(S\cup A)\geq k,
\]

since $S\cup A\neq A$ as $x\in S-A$ and there is at least one receiver
$y\notin(S\cup A)\cup T$ as $y\notin S\cup A,\, y\notin T.$ Thus
(\ref{eq:inequ_1-1-1}) implies

\[
\delta_{G-F}(S)\geq k
\]

and so,

\[
\delta_{G-F-e}(S)\geq k-1.
\]

Thus, we can increase $F$ till finally it will satisfy (i), (ii)
and reach all receivers $t_{i}$. Then apply the induction hypothesis
on $G-F$. This completes the proof.

$\qquad\qquad\qquad\qquad\qquad\qquad\qquad\qquad\qquad\qquad\qquad\qquad\blacksquare$

The obove proof yields an efficient algorithm to construct a maximum
set of edge-disjoint trees reaching all receivers. Let

\[
K(G)=\underset{S\subset V(G),\, a\in S,\,\exists t_{i}\in V(G)-S}{\min}\delta_{G}(S)
\]

These trees can be constructed edge by edge. At any stage, we can
increase $F$ by checking at most $E(G)$ edges $e$ whether or not

\[
K(G-F-e)\geq k-1.
\]

Since determining $K(G)$ can be done in $p$ steps, where $p$ is
a polynomial in $V(G)$, $E(G)$. Thus, we can obtain $k$ edge-disjoint
trees in at most $O(pE(G))$ steps.

Over the two layer graph $\mathcal{D}_{m}$, The algorithm packs unit-capacity
trees one by one. Each unit-capacity tree is constructed by greedily
constructing a tree edge by edge starting from the source and augmenting
towards all receivers. It is similar to the greedy treepacking algorithm
based on Prim's algorithm. The distinction lies in the rule of selecting
the edge among all potential edges. The edge whose removal leads to
least reduction in the multicast capacity of the residual graph is
chosen in the greedy algorithm.

Because we alway choose the edge whose removal leads to least reduction
in the multicast capacity of the residual graph, the edge we choose
can alway satisfy $K(G-F-e)\geq k-1$. Therefore, based on the above
proof, finally we can obtain $k$ edge-disjoint trees.

Due to the special structure of $\mathcal{D}_{m}$, the time complexity
of computing $K(G)$ is $O(V(G)*E(G))$. Therefore, the time complexity
of the algorithm is $O(V(G)*E^{2}(G)).$

$\qquad\qquad\qquad\qquad\qquad\qquad\qquad\qquad\qquad\qquad\qquad\qquad\blacksquare$

\textbf{Proof of the inequality (}\ref{eq:inequ_1-1-1}\textbf{)}.

\textit{Proof}: suppose $e=(x,\, y)\in\vartriangle_{G-F}(S\cup A),$
then $x\in S\cup A,$ and $y\in V(G)-S-A,$ thus we must have
\[
e\in\vartriangle_{G-F}(S)\cup\vartriangle_{G-F}(A).
\]
 Similarly, suppose $e=(x,\, y)\in\vartriangle_{G-F}(S\cap A),$ then
$x\in S\cap A,$ and $y\in V(G)-S\cap A,$ we also have

\[
e\in\vartriangle_{G-F}(S)\cup\vartriangle_{G-F}(A).
\]

if $e=(x,\, y)\in\vartriangle_{G-F}(S\cup A)\cap\vartriangle_{G-F}(S\cap A)$,
then $x\in S\cap A,$ and $y\in V(G)-S-A$. Therefore we have

\[
e\in\vartriangle_{G-F}(S)\cap\vartriangle_{G-F}(A).
\]

Base on the above observation, we can have

\[
\delta_{G-F}(S\cup A)+\delta_{G-F}(S\cap A)\leq\delta_{G-F}(S)+\delta_{G-F}(A)
\]

$\qquad\qquad\qquad\qquad\qquad\qquad\qquad\qquad\qquad\qquad\qquad\qquad\blacksquare$
\begin{description}
\item [{{{{{{{{{{{{B.}}}}}}}}}}}}] Proof of Corollary
$2$
\end{description}
\textit{Proof:} Let a length-$|E|$ binary vector $I_{X}$ be the
indicator vector for edge set $X\subseteq E$; its $e$-th entry is
1 if $e\epsilon X$, and 0 otherwise.

Since $R_{m}(\boldsymbol{c}_{m},D)$ is the minimum min-cut over $\mathcal{D}_{m}$.
Therefore it can be expressed as

\[
R_{m}(\boldsymbol{c}_{m},D)=\underset{i\epsilon T}{\min}\underset{U:\, s\epsilon U,\, t_{i}\epsilon\bar{U}}{\min}I_{\delta(U)}\boldsymbol{c}_{m}
\]

where $\delta(U)$ denote the set of edges going from $U$ to$\bar{U}$.
So $R_{m}(\boldsymbol{c}_{m},D)$ is the pointwise minimum of a family
of linear functions. Let $\boldsymbol{c}_{m}^{1}$ and $\boldsymbol{c}_{m}^{2}$
denote two different link rate vector, and $\lambda_{1}+\lambda_{2}=1,\,\lambda_{1}\geq0,\,\lambda_{2}\geq0$.

Then we have

\begin{eqnarray*}
R_{m}(\lambda_{1}\boldsymbol{c}_{m}^{1}+\lambda_{2}\boldsymbol{c}_{m}^{2},D) & = & \underset{i\epsilon T}{\min}\underset{U:\, s\epsilon U,\, t_{i}\epsilon\bar{U}}{\min}I_{\delta(U)}(\lambda_{1}\boldsymbol{c}_{m}^{1}+\lambda_{2}\boldsymbol{c}_{m}^{2})\\
 & \geq & \underset{i\epsilon T}{\min}\underset{U:\, s\epsilon U,\, t_{i}\epsilon\bar{U}}{\min}I_{\delta(U)}(\lambda_{1}\boldsymbol{c}_{m}^{1})\\
 &  & +\underset{i\epsilon T}{\min}\underset{U:\, s\epsilon U,\, t_{i}\epsilon\bar{U}}{\min}I_{\delta(U)}(\lambda_{2}\boldsymbol{c}_{m}^{2})\\
 & = & R_{m}(\lambda_{1}\boldsymbol{c}_{m}^{1},D)+R_{m}(\lambda_{2}\boldsymbol{c}_{m}^{2},D)\\
 & = & \lambda_{1}R_{m}(\boldsymbol{c}_{m}^{1},D)+\lambda_{2}R_{m}(\boldsymbol{c}_{m}^{2},D)
\end{eqnarray*}

So $R_{m}(\boldsymbol{c}_{m},D)$ is a concave function of the overlay
link rates $\boldsymbol{c}_{m}$.

$\qquad\qquad\qquad\qquad\qquad\qquad\qquad\qquad\qquad\qquad\qquad\qquad\blacksquare$
\begin{description}
\item [{{{{{{{{{{{{C.}}}}}}}}}}}}] Proof of Proposition
$1$
\end{description}
$Proof:$ For $any\, e\epsilon E$ and $m=1,...\, M$, let $c_{m,e}^{(1)}$
and $c_{m,e}^{(2)}$ denote two differnet value. It is easy to verified
that $-\sum_{l\in\mathcal{L}}\int_{0}^{\boldsymbol{a}_{l}^{T}\boldsymbol{c}}\frac{(z-C_{l})^{+}}{z}\, dz$
is a concave function and $-\sum_{l\in\mathcal{L}}a_{l,e}\frac{(\boldsymbol{a}_{l}^{T}\boldsymbol{y}-C_{l})^{+}}{\boldsymbol{a}_{l}^{T}\boldsymbol{y}}$
is its subgradient with respect to $c_{m,e}$. Therefore, we just
need to show $U'_{m}\left(R_{m}\right)\frac{\partial R_{m}}{\partial c_{m,e}}$
is a subgradient of $U{}_{m}\left(R_{m}\right)$ with respect to $c_{m,e}$.

Since $U{}_{m}\left(R_{m}\right)$ is a increasing and strictly concave
function and $R_{m}(\boldsymbol{c}_{m},D)$ is a concave function
with respect to $\boldsymbol{c}_{m}$, which has been proved in Corollary
1. Then we can have

\[
U{}_{m}\left(R_{m}^{(1)}\right)-U{}_{m}\left(R_{m}^{(2)}\right)\leq U'_{m}\left(R_{m}^{(2)}\right)(R_{m}^{(1)}-R_{m}^{(2)})
\]

\[
R_{m}^{(1)}-R_{m}^{(2)}\leq\frac{\partial R_{m}^{(2)}}{\partial c_{m,e}}(c_{m,e}^{(1)}-c_{m,e}^{(2)})
\]

Since$U{}_{m}\left(R_{m}\right)$ is nondecreasing, we have $U'_{m}\left(R_{m}^{(2)}\right)\geq0$.
Then

\begin{eqnarray*}
U{}_{m}\left(R_{m}^{(1)}\right)-U{}_{m}\left(R_{m}^{(2)}\right) & \leq & U'_{m}\left(R_{m}^{(2)}\right)(R_{m}^{(1)}-R_{m}^{(2)})\\
 & \leq & U'_{m}\left(R_{m}^{(2)}\right)\frac{\partial R_{m}^{(2)}}{\partial c_{m,e}}(c_{m,e}^{(1)}-c_{m,e}^{(2)})
\end{eqnarray*}

Therefore, $U'_{m}\left(R_{m}\right)\frac{\partial R_{m}}{\partial c_{m,e}}$
is a subgradient of $U{}_{m}\left(R_{m}\right)$ with respect to $c_{m,e}$.

$\qquad\qquad\qquad\qquad\qquad\qquad\qquad\qquad\qquad\qquad\qquad\qquad\blacksquare$
\begin{description}
\item [{{{{D.}}}}] Proof of Theorem $3$
\end{description}
\textit{Proof:} Let %
{} %
$g=\underset{l\epsilon\mathcal{L}}{\max}\frac{1}{C_{l}}$, $A$=$diag(C_{l},\, l\epsilon\mathcal{L})$.
let $(\boldsymbol{c}^{*},\,\boldsymbol{p}^{*})$ be a saddle point
of the Lagrangian function $\mathcal{G}\left(\boldsymbol{c},\boldsymbol{p}\right)$.
We use $\mathcal{G}_{\boldsymbol{c}}\left(\boldsymbol{c},\boldsymbol{p}\right)$and
$\mathcal{G}_{\boldsymbol{p}}\left(\boldsymbol{c},\boldsymbol{p}\right)$
to denote a subgradient of $\mathcal{G}\left(\boldsymbol{c},\boldsymbol{p}\right)$
with respect to $\boldsymbol{c}$ and a subgradient of $\mathcal{G}\left(\boldsymbol{c},\boldsymbol{p}\right)$with
respect to $\boldsymbol{p}$. Suppose that $|U'_{m}(R_{m}(\boldsymbol{c}_{m}))|$
$\forall m\epsilon M$ is upper bounded by a positive constant $\bar{U}$.

Under the assumption that $|U'_{m}(R_{m}(\boldsymbol{c}_{m}))|$ $\forall m\epsilon M$
is upper bounded by a positive constant $\bar{U}$, there is a constant
$\triangle>0$ , such that $||\mathcal{G}_{\boldsymbol{c}}\left(\boldsymbol{c}^{(k)},\boldsymbol{p}^{(k)}\right)||_{2}\leq\triangle$,
and $||\mathcal{G}_{\boldsymbol{p}}\left(\boldsymbol{c}^{(k)},\boldsymbol{p}^{(k)}\right)||_{2}\leq\triangle$
for all $k\geq0$.

In order to prove theorem 2, we need to prove the following two lemmas.

\textbf{Lemma 1}: (a) For any $\boldsymbol{c}\geq0$ and all $k\geq0$,

\begin{eqnarray*}
||\boldsymbol{c}^{(k+1)}-\boldsymbol{c}||_{2}^{2} & \leq & ||\boldsymbol{c}^{(k)}-\boldsymbol{c}||_{2}^{2}+2\alpha\left[\mathcal{G}\left(\boldsymbol{c}^{(k)},\boldsymbol{p}^{(k)}\right)\right.\\
 &  & \left.-\mathcal{G}\left(\boldsymbol{c},\boldsymbol{p}^{(k)}\right)\right]+\alpha^{2}||\mathcal{G}_{\boldsymbol{c}}\left(\boldsymbol{c}^{(k)},\boldsymbol{p}^{(k)}\right)||_{2}^{2}
\end{eqnarray*}

(b)For any $\boldsymbol{p}\geq0$ and all $k\geq0$,

\begin{eqnarray*}
\left(\boldsymbol{p}^{(k+1)}-\boldsymbol{p}\right)^{T}A\left(\boldsymbol{p}^{(k+1)}-\boldsymbol{p}\right) & \leq & \left(\boldsymbol{p}^{(k)}-\boldsymbol{p}\right)^{T}A\left(\boldsymbol{p}^{(k)}-\boldsymbol{p}\right)\\
 &  & -2\left[\mathcal{G}\left(\boldsymbol{c}^{(k)},\boldsymbol{p}^{(k)}\right)-\mathcal{G}\left(\boldsymbol{c}^{(k)},\boldsymbol{p}\right)\right]\\
 &  & +g||\mathcal{G}_{\boldsymbol{p}}\left(\boldsymbol{c}^{(k)},\boldsymbol{p}^{(k)}\right)||_{2}^{2}
\end{eqnarray*}

\textit{Proof:} (a) From the algorithm (9)-(10), we obtain that for
any $\boldsymbol{c}\geq0$ and all $k>0$,

\begin{eqnarray*}
||\boldsymbol{c}^{(k+1)}-\boldsymbol{c}||_{2}^{2} & \leq & ||\boldsymbol{c}^{(k)}+a\mathcal{G}_{\boldsymbol{c}}\left(\boldsymbol{c}^{(k)},\boldsymbol{p}^{(k)}\right)-\boldsymbol{c}||_{2}^{2}\\
 & = & ||\boldsymbol{c}^{(k)}-\boldsymbol{c}||_{2}^{2}+2\alpha\mathcal{G}_{\boldsymbol{c}}\left(\boldsymbol{c}^{(k)},\boldsymbol{p}^{(k)}\right)^{T}\left(\boldsymbol{c}^{(k)}-\boldsymbol{c}\right)\\
 &  & +\alpha^{2}||\mathcal{G}_{\boldsymbol{c}}\left(\boldsymbol{c}^{(k)},\boldsymbol{p}^{(k)}\right)||_{2}^{2}
\end{eqnarray*}

Since the function $\mathcal{G}\left(\boldsymbol{c},\boldsymbol{p}\right)$
is concave in $\boldsymbol{c}$ for each $\boldsymbol{p}\geq0$, and
since $\mathcal{G}_{\boldsymbol{c}}\left(\boldsymbol{c}^{(k)},\boldsymbol{p}^{(k)}\right)$
is a subgradient of $\mathcal{G}\left(\boldsymbol{c},\boldsymbol{p}^{(k)}\right)$
with respect to $\boldsymbol{c}$ at $\boldsymbol{c}=\boldsymbol{c}^{(k)}$,
we obtain for any $\boldsymbol{c}$,

\[
\mathcal{G}_{\boldsymbol{c}}\left(\boldsymbol{c}^{(k)},\boldsymbol{p}^{(k)}\right)^{T}\left(\boldsymbol{c}^{(k)}-\boldsymbol{c}\right)\leq\mathcal{G}\left(\boldsymbol{c}^{(k)},\boldsymbol{p}^{(k)}\right)-\mathcal{G}\left(\boldsymbol{c},\boldsymbol{p}^{(k)}\right)
\]

Hence, for any $\boldsymbol{c}\geq0$ and all $k\geq0$,

\begin{eqnarray*}
||\boldsymbol{c}^{(k+1)}-\boldsymbol{c}||_{2}^{2} & \leq & ||\boldsymbol{c}^{(k)}-\boldsymbol{c}||_{2}^{2}+2\alpha\left[\mathcal{G}\left(\boldsymbol{c}^{(k)},\boldsymbol{p}^{(k)}\right)\right.\\
 &  & \left.-\mathcal{G}\left(\boldsymbol{c},\boldsymbol{p}^{(k)}\right)\right]+\alpha^{2}||\mathcal{G}_{\boldsymbol{c}}\left(\boldsymbol{c}^{(k)},\boldsymbol{p}^{(k)}\right)||_{2}^{2}
\end{eqnarray*}

(b) Similarly, from (9)-(10), for any $p_{l}\geq0$ $l\epsilon\mathcal{L}$,
we have,

\begin{eqnarray*}
C_{l}|p_{l}^{(k+1)}-p_{l}|^{2} & \leq & C_{l}|p_{l}^{(k)}-p_{l}|^{2}-2\left(p^{(k)}-p_{l}\right)\mathcal{G}_{p_{l}}\left(\boldsymbol{c}^{(k)},\boldsymbol{p}^{(k)}\right)\\
 &  & +\frac{1}{C_{l}}|\mathcal{G}_{p_{l}}\left(\boldsymbol{c}^{(k)},\boldsymbol{p}^{(k)}\right)|^{2}
\end{eqnarray*}

By adding these relations over all $l\epsilon\mathcal{L}$. we obtain
for any $\boldsymbol{p}\geq0$ and all $k\geq0$.

\begin{eqnarray*}
\left(\boldsymbol{p}^{(k+1)}-\boldsymbol{p}\right)^{T}A\left(\boldsymbol{p}^{(k+1)}-\boldsymbol{p}\right) & \leq & \left(\boldsymbol{p}^{(k)}-\boldsymbol{p}\right)^{T}A\left(\boldsymbol{p}^{(k)}-\boldsymbol{p}\right)\\
 &  & -2\left(\boldsymbol{p}^{(k)}-\boldsymbol{p}\right)^{T}\mathcal{G}_{\boldsymbol{p}}\left(\boldsymbol{c}^{(k)},\boldsymbol{p}^{(k)}\right)\\
 &  & +g||\mathcal{G}_{\boldsymbol{p}}\left(\boldsymbol{c}^{(k)},\boldsymbol{p}^{(k)}\right)||_{2}^{2}
\end{eqnarray*}

Since $\mathcal{G}_{\boldsymbol{p}}\left(\boldsymbol{c}^{(k)},\boldsymbol{p}^{(k)}\right)$
is a subgradient of the linear function $\mathcal{G}\left(\boldsymbol{c}^{(k)},\boldsymbol{p}\right)$
at $\boldsymbol{p}=\boldsymbol{p}^{(k)}$, we have for all $\boldsymbol{p}$.

\[
\left(\boldsymbol{p}^{(k)}-\boldsymbol{p}\right)^{T}\mathcal{G}_{\boldsymbol{p}}\left(\boldsymbol{c}^{(k)},\boldsymbol{p}^{(k)}\right)=\mathcal{G}\left(\boldsymbol{c}^{(k)},\boldsymbol{p}^{(k)}\right)-\mathcal{G}\left(\boldsymbol{c}^{(k)},\boldsymbol{p}\right)
\]

Therefore for any $\boldsymbol{p}\geq0$ and all $k>0$.

\begin{eqnarray*}
\left(\boldsymbol{p}^{(k+1)}-\boldsymbol{p}\right)^{T}A\left(\boldsymbol{p}^{(k+1)}-\boldsymbol{p}\right) & \leq & \left(\boldsymbol{p}^{(k)}-\boldsymbol{p}\right)^{T}A\left(\boldsymbol{p}^{(k)}-\boldsymbol{p}\right)\\
 &  & -2\left[\mathcal{G}\left(\boldsymbol{c}^{(k)},\boldsymbol{p}^{(k)}\right)-\mathcal{G}\left(\boldsymbol{c}^{(k)},\boldsymbol{p}\right)\right]\\
 &  & +g||\mathcal{G}_{\boldsymbol{p}}\left(\boldsymbol{c}^{(k)},\boldsymbol{p}^{(k)}\right)||_{2}^{2}
\end{eqnarray*}

$\qquad\qquad\qquad\qquad\qquad\qquad\qquad\qquad\qquad\qquad\qquad\qquad\blacksquare$

\textbf{Lemma 2:} let $\hat{\boldsymbol{c}(k)}$ and $\hat{\boldsymbol{p}(k)}$
be the iterate averages given by

\[
\hat{\boldsymbol{c}}(k)=\frac{1}{k}\overset{k-1}{\underset{i=0}{\sum}}\boldsymbol{c}^{(i)},\qquad\hat{\boldsymbol{p}}(k)=\frac{1}{k}\overset{k-1}{\underset{i=0}{\sum}}\boldsymbol{p}^{(i)}.
\]

we then have for all $k\geq1$,

\begin{equation}
\frac{-1}{2\alpha k}||\boldsymbol{c}^{(0)}-\boldsymbol{c}||_{2}^{2}-\frac{\alpha\triangle^{2}}{2}\leq\frac{1}{k}\overset{k-1}{\underset{i=0}{\sum}}\mathcal{G}\left(\boldsymbol{c}^{(i)},\boldsymbol{p}^{(i)}\right)-\mathcal{G}\left(\boldsymbol{c},\hat{\boldsymbol{p}}(k)\right)\label{eq:lem2_1}
\end{equation}

\begin{equation}
\frac{1}{k}\overset{k-1}{\underset{i=0}{\sum}}\mathcal{G}\left(\boldsymbol{c}^{(i)},\boldsymbol{p}^{(i)}\right)-\mathcal{G}\left(\hat{\boldsymbol{c}}(k),\boldsymbol{p}\right)\leq\frac{g\triangle^{2}}{2}+\frac{\left(\boldsymbol{p}^{(0)}-\boldsymbol{p}\right)^{T}A\left(\boldsymbol{p}^{(0)}-\boldsymbol{p}\right)}{2k}\label{eq:lem2_2}
\end{equation}

\textit{Proof:} by using Corollary 1 and Lemma 1(a), we have for any
$\boldsymbol{c}\geq0$ and $i\geq0$,

\begin{eqnarray*}
\frac{1}{2\alpha}\left[||\boldsymbol{c}^{(i+1)}-\boldsymbol{c}||_{2}^{2}-||\boldsymbol{c}^{(i)}-\boldsymbol{c}||_{2}^{2}\right]-\frac{\alpha}{2}\triangle^{2} & \leq & \mathcal{G}\left(\boldsymbol{c}^{(i)},\boldsymbol{p}^{(i)}\right)\\
 &  & -\mathcal{G}\left(\boldsymbol{c},\boldsymbol{p}^{(i)}\right)
\end{eqnarray*}

By adding these relations over $i=0,...,k-1$, we obtain for any $\boldsymbol{c}\geq0$
and $k\geq1$,

\begin{eqnarray*}
-\frac{1}{2k\alpha}||\boldsymbol{c}^{(0)}-\boldsymbol{c}||_{2}^{2}-\frac{\alpha}{2}\triangle^{2}\\
\leq\frac{1}{2k\alpha}\left[||\boldsymbol{c}^{(k)}-\boldsymbol{c}||_{2}^{2}-||\boldsymbol{c}^{(0)}-\boldsymbol{c}||_{2}^{2}\right]-\frac{\alpha}{2}\triangle^{2}\\
\leq\frac{1}{k}\overset{k-1}{\underset{i=0}{\sum}}\left[\mathcal{G}\left(\boldsymbol{c}^{(i)},\boldsymbol{p}^{(i)}\right)-\mathcal{G}\left(\boldsymbol{c},\boldsymbol{p}^{(i)}\right)\right]
\end{eqnarray*}

Since the function $\mathcal{G}\left(\boldsymbol{c},\boldsymbol{p}\right)$
is linear in $\boldsymbol{p}$ for any fixed $\boldsymbol{c}\geq0$,
there holds

\[
\mathcal{G}\left(\boldsymbol{c},\hat{\boldsymbol{p}}(k)\right)=\frac{1}{k}\overset{k-1}{\underset{i=0}{\sum}}\mathcal{G}\left(\boldsymbol{c},\boldsymbol{p}^{(i)}\right)
\]

Combining the preceding two relations, we obtain for any $\boldsymbol{c}\geq0$
and $k\geq1$,

\[
-\frac{1}{2k\alpha}||\boldsymbol{c}^{(0)}-\boldsymbol{c}||_{2}^{2}-\frac{\alpha}{2}\triangle^{2}\leq\frac{1}{k}\overset{k-1}{\underset{i=0}{\sum}}\mathcal{G}\left(\boldsymbol{c}^{(i)},\boldsymbol{p}^{(i)}\right)-\mathcal{G}\left(\boldsymbol{c},\hat{\boldsymbol{p}}(k)\right)
\]

thus establishing relation (\ref{eq:lem2_1}).

Similarly, by using Corollary 1 and Lemma 1(b),we have for any $\boldsymbol{p}\geq0$
and $i\geq0$,

\begin{eqnarray*}
\left(\boldsymbol{p}^{(i+1)}-\boldsymbol{p}\right)^{T}A\left(\boldsymbol{p}^{(i+1)}-\boldsymbol{p}\right) & \leq & \left(\boldsymbol{p}^{(i)}-\boldsymbol{p}\right)^{T}A\left(\boldsymbol{p}^{(i)}-\boldsymbol{p}\right)\\
 &  & -2\left[\mathcal{G}\left(\boldsymbol{c}^{(i)},\boldsymbol{p}^{(i)}\right)\right.\\
 &  & \left.-\mathcal{G}\left(\boldsymbol{c}^{(i)},\boldsymbol{p}\right)\right]+g\triangle^{2}
\end{eqnarray*}

By adding these relations over $i=0,...,k-1$, we obtain for any $\boldsymbol{p}\geq0$
and $k\geq1$,

\begin{eqnarray*}
\frac{1}{k}\overset{k-1}{\underset{i=0}{\sum}}\left[\mathcal{G}\left(\boldsymbol{c}^{(i)},\boldsymbol{p}^{(i)}\right)-\mathcal{G}\left(\boldsymbol{c}^{(i)},\boldsymbol{p}\right)\right]-\frac{g\triangle^{2}}{2}\\
\leq\frac{\left(\boldsymbol{p}^{(0)}-\boldsymbol{p}\right)^{T}A\left(\boldsymbol{p}^{(0)}-\boldsymbol{p}\right)}{2k}-\frac{\left(\boldsymbol{p}^{(k)}-\boldsymbol{p}\right)^{T}A\left(\boldsymbol{p}^{(k)}-\boldsymbol{p}\right)}{2k}\\
\leq\frac{\left(\boldsymbol{p}^{(0)}-\boldsymbol{p}\right)^{T}A\left(\boldsymbol{p}^{(0)}-\boldsymbol{p}\right)}{2k}
\end{eqnarray*}

because the function $\mathcal{G}\left(\boldsymbol{c},\boldsymbol{p}\right)$
is concave in $\boldsymbol{c}$ for any fixed $\boldsymbol{p}\geq0$,
we have

\[
\frac{1}{k}\overset{k-1}{\underset{i=0}{\sum}}\mathcal{G}\left(\boldsymbol{c}^{(i)},\boldsymbol{p}\right)\leq\mathcal{G}\left(\hat{\boldsymbol{c}}(k),\boldsymbol{p}\right)
\]

Combining the preceding two relations, we obtain for any $\boldsymbol{p}\geq0$
and $k\geq1$,

\begin{eqnarray*}
\frac{1}{k}\overset{k-1}{\underset{i=0}{\sum}}\mathcal{G}\left(\boldsymbol{c}^{(i)},\boldsymbol{p}^{(i)}\right)-\mathcal{G}\left(\hat{\boldsymbol{c}}(k),\boldsymbol{p}\right) & \leq & \frac{g\triangle^{2}}{2}+\frac{\left(\boldsymbol{p}^{(0)}-\boldsymbol{p}\right)^{T}A\left(\boldsymbol{p}^{(0)}-\boldsymbol{p}\right)}{2k}
\end{eqnarray*}

$\qquad\qquad\qquad\qquad\qquad\qquad\qquad\qquad\qquad\qquad\qquad\qquad\blacksquare$

Our proof of this theorem is based on Lemma 2. In particular, by letting
$\boldsymbol{c}=\boldsymbol{c}^{*}$ and $\boldsymbol{p}=\boldsymbol{p}^{*}$
in equations (\ref{eq:lem2_1}) and (\ref{eq:lem2_2}), repectively,
we obtain,

\[
\frac{-1}{2\alpha k}||\boldsymbol{c}^{(0)}-\boldsymbol{c}^{*}||_{2}^{2}-\frac{\alpha\triangle^{2}}{2}\leq\frac{1}{k}\overset{k-1}{\underset{i=0}{\sum}}\mathcal{G}\left(\boldsymbol{c}^{(i)},\boldsymbol{p}^{(i)}\right)-\mathcal{G}\left(\boldsymbol{c}^{*},\hat{\boldsymbol{p}}(k)\right)
\]

\[
\frac{1}{k}\overset{k-1}{\underset{i=0}{\sum}}\mathcal{G}\left(\boldsymbol{c}^{(i)},\boldsymbol{p}^{(i)}\right)-\mathcal{G}\left(\hat{\boldsymbol{c}}(k),\boldsymbol{p}^{*}\right)\leq\frac{g\triangle^{2}}{2}+\frac{\left(\boldsymbol{p}^{(0)}-\boldsymbol{p}^{*}\right)^{T}A\left(\boldsymbol{p}^{(0)}-\boldsymbol{p}^{*}\right)}{2k}
\]

By the saddle-point relation, we have

\[
\mathcal{G}\left(\hat{\boldsymbol{c}}(k),\boldsymbol{p}^{*}\right)\leq\mathcal{G}\left(\boldsymbol{c}^{*},\boldsymbol{p}^{*}\right)\leq\mathcal{G}\left(\boldsymbol{c}^{*},\hat{\boldsymbol{p}}(k)\right)
\]

Combining the preceding three relations, we obtain for all $k\geq1$,

\begin{eqnarray*}
\frac{-1}{2\alpha k}||\boldsymbol{c}^{(0)}-\boldsymbol{c}^{*}||_{2}^{2}-\frac{\alpha\triangle^{2}}{2} & \leq & \overset{k-1}{\underset{i=0}{\sum}}\mathcal{G}\left(\boldsymbol{c}^{(i)},\boldsymbol{p}^{(i)}\right)-\mathcal{G}\left(\boldsymbol{c}^{*},\boldsymbol{p}^{*}\right)\\
 & \leq & \frac{g\triangle^{2}}{2}+\frac{\left(\boldsymbol{p}^{(0)}-\boldsymbol{p}^{*}\right)^{T}A\left(\boldsymbol{p}^{(0)}-\boldsymbol{p}^{*}\right)}{2k}
\end{eqnarray*}

$\qquad\qquad\qquad\qquad\qquad\qquad\qquad\qquad\qquad\qquad\qquad\qquad\blacksquare$

\begin{algorithm}[h]
/{*}

Every 200ms each peer measures the loss rate and queuing delay of
its incoming links and gets the source sending rate from the packets
of corresponding session source it has received and adjusts the rates
of these links based on the link rate control algorithm, and then
sends them to their corresponding upstream senders for the new rates
to take eff{}ect.

$S$ denotes the set of all sessions. $s_{m}$ denotes the session
of peer $m$.

$E_{m}$ denotes the set of incoming links of peer $m$. $I_{m,e}$
is the critical link indicator of link $e$ for session $s_{m}$.
If $e$ is a critical link, then $I_{m,e}=1$, otherwise, $I_{m,e}=0$.

{*}/

1: $\textbf{for all}$ $e\epsilon E_{m}$ $\textbf{do}$

$\quad\qquad$/{*}get the loss rate of the link $e$ {*}/

2: $\qquad$$lossrate$$\leftarrow$GetAverageLoss();

$\quad\qquad$/{*}get the queuing delay of the link $e$ {*}/

3: $\qquad$$queuing_{-}delay$$\leftarrow$GetAverageQueuingDelay();

4: $\qquad$$\textbf{for all}$ $s\epsilon S$ $\textbf{do}$

5: $\qquad$$\qquad$$\textbf{if}$ $s\neq s_{m}$ $\textbf{then}$

$\quad\qquad$$\qquad$$\qquad$/{*} get the source sending rate of
session $s$ {*}/

6: $\qquad$$\qquad$$\qquad$$sending_{-}rate$$\leftarrow$GetSourceSendingRate();

$\quad\qquad$$\qquad$$\qquad$ /{*} get the critical cut indicator
of link $e$

$\quad\qquad$$\qquad$$\qquad$for session $s$ {*}/

7: $\qquad$$\qquad$$\qquad$$I_{m,e}$$\leftarrow$GetCriticalCut($e$
,$m$);

8: $\qquad$$\qquad$$\qquad$$delta$$\leftarrow$$step_{-}size$($\beta/$$sending_{-}rate$

$\quad\qquad$$\qquad$$\qquad$-$lossrate$-$queuing_{-}delay$);

9: $\qquad$$\qquad$$\qquad$$list$.push\_back(pair<$s$, $delta$>);

10:$\qquad$$\qquad$$\textbf{end if}$

11:$\qquad$$\textbf{end for}$

$\quad\qquad$ /{*} send the updated rate to the upstream of the link
{*}/

12:$\qquad$Update($e$, $list$);

13:$\textbf{end for}$

\caption{Link Rate Control}
\end{algorithm}

\begin{algorithm}[h]
 /{*}

Every 300ms each source peer packs trees using the link states it
collects and calculates the critical cut information, and then append
the critical cut information and source sending rate in the header
of the packets that it will send out through these trees.

$S$ denotes the set of all sessions. $s_{m}$ denotes the session
of peer $m$.

$Link_{-}States_{m}$ is the collected links states for session $s_{m}$.

{*}/

/{*} source peer $m$ packs delay-limited trees. {*}/

1: $Trees\leftarrow$PackTree($Link_{-}States_{m})$

/{*} calculate the critical cut information for session $s_{m}$

2: $I_{m}\leftarrow$CalculateCriticalCut($Link_{-}States_{m})$;

/{*} deliver packet {*}/

3: $\textbf{while}$ (CanSendPacket()) $\textbf{do}$

$\quad\qquad$/{*} get a tree with the maximum rate among the trees
{*}/

4: $\qquad$$tree$$\leftarrow$GetATree($Trees$);

5: $\qquad$$datapacket$$\leftarrow$CreatePacket();

6: $\qquad$$sending_{-}rate$$\leftarrow0$;

7: $\qquad$$\textbf{for all}$$t\epsilon Trees$ $\textbf{do}$

8: $\qquad$$\qquad$$sending_{-}rate\leftarrow sending_{-}rate+t.rate$;

9: $\qquad$$\textbf{end for}$

$\quad\qquad$/{*} add the critical cut information and source\_sending\_rate

$\quad\qquad$to the header of the packet {*}/

10:$\qquad$Append($datapacket$, $I_{m}$, $sending_{-}rate$);

11:$\qquad$Deliver($datapacket,tree$);

12:$\textbf{end while}$\caption{Data Multicast}
\end{algorithm}

\bibliographystyle{IEEEtran}
\bibliography{IEEEabrv,ref}


\end{document}